\documentclass[prb, aps, showpacs,reprint]{revtex4-1}

\usepackage[utf8]{inputenc}
\usepackage{graphicx}
\usepackage{amsmath}
\usepackage{amssymb}
\usepackage{braket}
\usepackage{color}

\renewcommand\vec[1]{\ensuremath\boldsymbol{#1}}
\newcommand{\uvec}[1]{\hat{\mathbf{#1}}}

\newcommand{\diff}[1]{\mathrm{d} #1}

\newcommand{\spinup}{\uparrow}
\newcommand{\spindown}{\downarrow}

\newcommand{\dtime}[1]{\dot{#1}}

\newcommand{\nearnb}[2]{\left< #1 , #2 \right>}
\newcommand{\Sec}[0]{2}
\newcommand{\Ins}[0]{\textnormal{AF}}
\newcommand{\Int}[0]{I}
\newcommand{\Cond}[0]{N}
\newcommand{\Tot}[0]{T}
\newcommand{\Th}[0]{Q}

\newcommand{\FD}[0]{n_{F}}
\newcommand{\BE}[0]{n_{B}}
\newcommand{\kB}[0]{k_{B}}
\newcommand{\FE}[0]{F}
\newcommand{\Neel}[0]{\text{Neel}}
\newcommand{\Fin}[0]{f}
\newcommand{\Ini}[0]{i}

\newcommand{\diffv}[1]{\mathrm{d}^3 #1}

\newcommand{\U}[0]{U}

\newcommand{\R}[0]{R}
\newcommand{\A}[0]{d}
\newcommand{\D}[0]{d_{I}}
\newcommand{\xN}[0]{x_{I}}

\DeclareMathOperator{\Tr}{Tr}

\DeclareMathOperator{\acosh}{arccosh}

\providecommand{\abs}[1]{\lvert#1\rvert} 

\begin{document}

\title{Electrically Driven Bose-Einstein Condensation of Magnons in Antiferromagnets}

\author{Eirik Løhaugen Fjærbu}
\author{Niklas Rohling}
\author{Arne Brataas}
\affiliation{Department of Physics, Norwegian University of Science and Technology, NO-7491, Trondheim, Norway}

\date{\today}


\begin{abstract}
We explore routes to realize electrically driven Bose-Einstein condensation of magnons in insulating antiferromagnets. 
Even in insulating antiferromagnets, the localized spins can strongly couple to itinerant spins in adjacent metals via spin-transfer torque and spin pumping. 
We describe the formation of steady-state magnon condensates controlled by a spin accumulation polarized along the staggered field in an adjacent normal metal. 
Two types of magnons exist in antiferromagnets, which carry opposite magnetic moments. Consequently, and in contrast to ferromagnets, Bose-Einstein condensation can occur for either sign of the spin accumulation. 
This condensation may occur even at room temperature when the interaction with the normal metal is fast compared to the relaxation processes within the antiferromagnet. 
In antiferromagnets, the operating frequencies of the condensate are orders of magnitude faster than in ferromagnets.  
\end{abstract}

\maketitle

\section{\label{sec:intro}Introduction}

Bose-Einstein condensation (BEC) occurs in a wide variety of systems\cite{Dalfovo:rmp1999, Nikuni:prl2000, Ruegg:nature2003, Radu:prl2005, Demokritov:Nature2006, Kasprzak:nature2006, Balili:science2007, Klaers:nature2010, Bunkov:JETP2011}. 
At a sufficient density, magnons condense into a single Bose quantum state. 
Spectroscopically generated magnon condensates have been observed in ferrimagnetic insulators at room temperature\cite{Demokritov:Nature2006}. 
These findings imply that it is feasible to demonstrate coherent quantum phenomena using magnons.
These effects could potentially be used in devices without the need for complicated cooling equipment. 
Magnon BEC manifests itself through a phase-coherent precession of the magnetization and an accompanying peak in the population of the magnons at the lowest energy spin-wave mode. 
Associated with magnon BEC is the possibility of realizing and controlling spin superfluidity\cite{Sonin:ssc1978,Sonin:adp2010,Bunkov:prl2012,Chen:prb2014,Takei:prl2014,Takei:prb2014,Skarsvag:prl2015,Flebus:prl2016,Takei:prl2016,Sun:prl2016}. 
The superfluid properties could enable long-range dissipationless spin transport. 
In antiferromagnets, there are also reports of condensation and superfluidity induced by nuclear magnetic resonance\cite{Bunkov:JETP2011,Bunkov:prl2012}.

In this work, we explore an electrical route for controlling the Bose-Einstein condensation of magnons in antiferromagnetic insulators (AFIs). 
Spin pumping may be as operative from AFIs as from ferromagnetic insulators\cite{Cheng:prl2014}, in apparent contrast to naïve intuition. 
This means that the dynamical precession of spins in AFIs may pump pure spin currents into adjacent normal metals that are as large as those of ferromagnetic insulators. 
The effectiveness of spin pumping from antiferromagnetic insulators to metals implies, via Onsager reciprocity relations, that there is a considerable spin-transfer torque on the AFIs via spin accumulations in neighboring conductors\cite{Cheng:prl2014}. 
A spin accumulation can be generated via the spin Hall effect or from other ferromagnets. 
The combination of significant spin-transfer torques and spin pumping enables terahertz antiferromagnetic spin Hall nano-oscillators\cite{Cheng:prl2016}. 
For these reasons, antiferromagnetic insulators may be as effective as ferromagnetic insulators in spintronics devices.

There is currently considerable interest in coupling the electronic properties of normal metals and ferromagnetic insulators.
Although there is increasing attention on antiferromagnetic insulators\cite{Tveten:prl2013, Cheng:prl2014, Tveten:prl2014, Wang:prl2014, Takei:prb2014, Wang:prb2015, Takei:prb2015, Moriyama:apl2015, Cheng:prl2016, Tveten:prb2016}, they remain much less explored for spintronics purposes than their ferromagnetic counterparts. 
Whereas there are predictions of electrically driven magnon condensation in ferromagnets\cite{PhysRevLett.108.246601,bender2014dynamic}, the case of antiferromagnets is entirely unexplored.

In antiferromagnetic materials, the magnetic moments of the atoms exhibit a staggered (N\'eel) order, which gives rise to long-range correlations between the moments. 
However, the order is such that the net magnetization vanishes in each unit cell. 
An interesting aspect of antiferromagnets is that the spin dynamics can be a thousand times faster than the magnetization dynamics in ferromagnetic systems. 
Combining an antiferromagnetic insulator with a normal metal paves the way toward technological magnetic devices that operate at terahertz frequencies. 

To determine the feasibility of Bose-Einstein condensation of magnons in antiferromagnetic insulators, we generalize the theories of (staggered) spin transfer and spin pumping in normal metal-antiferromagnetic systems into the quantum domain with quantized spin wave excitations. 
To this end, we employ a quantum-mechanical model to describe both the localized electrons in the antiferromagnet and the conduction electrons in the metal. 
In the metal, a spin accumulation is assumed to exist via either the spin Hall effect or spin injection from additional ferromagnets. 
We use a description in which the spins in the AFI are exchange coupled to the itinerant spins in the metal. 
By computing the rates of change of the occupation of the magnons, we will determine the conditions for Bose-Einstein condensation driven by the spin accumulation. 
We will consider a quasi-equilibrated system, where the magnon lifetime is long compared to the thermalization time scale.

We consider easy-axis antiferromagnets. 
In such antiferromagnets, there are two types of magnons. 
The difference between these magnons is that they carry magnetic moments in opposite directions along the easy axis. 
The two types of magnons have identical dispersions and will therefore be equally occupied at equilibrium. 
Out-of-equilibrium, the two types of magnons are affected differently by the spin accumulation. 
Consequently, we will show that magnon condensation occurs for either polarity of the spin accumulation. 
This distinct feature in antiferromagnets is because the spin-transfer torque dampens the excitations of magnons of one type but can dramatically change the state of the other type\cite{Gomonay:prb2010,Cheng:prl2014,Cheng:prl2016}.

The remainder of this paper is organized as follows. 
We introduce the Hamiltonian describing the magnons in the antiferromagnet, the itinerant electrons in the normal metal, and the electron-magnon interaction across the AFI-normal metal interface in Sec.~\ref{sec:dynamics}. 
Next, we compute the transport rate for the transfer of spin angular momentum across the interface in Sec.~\ref{sec:rates}. 
In Sec.~\ref{sec:fixedT}, we compute the conditions for Bose-Einstein condensation in systems where the magnon temperature is held fixed by another reservoir. 
We conclude our paper in Sec.~\ref{sec:conclusions}. 
The Appendices \ref{sec:scattering.magnons}, \ref{sec:scattering.conduction}, and \ref{sec:scattering.coefficients} contain a substantial part of our work, including the microscopic calculation of the electron-magnon scattering amplitudes.

\section{\label{sec:dynamics}Dynamics}

The system consists of a normal metal, an antiferromagnetic insulator, and a metal-insulator interface. 
Across the interface, the itinerant spins in the normal metal are exchange coupled to the localized spins in the antiferromagnetic insulator. 
The system is illustrated in Fig.\ \ref{fig:intro.system}. 
In the normal metal, the electrons are driven out of equilibrium, thereby resulting in a spin accumulation. 
We consider a scenario in which there is a spin accumulation that is polarized along the $z$ axis, which is the easy axis of the antiferromagnet. 
The spin accumulation is induced by the spin Hall effect or by spin injection from ferromagnets\cite{RevModPhys.87.1213}. 

\subsection{Properties of the antiferromagnet}
\begin{figure}[ht]
	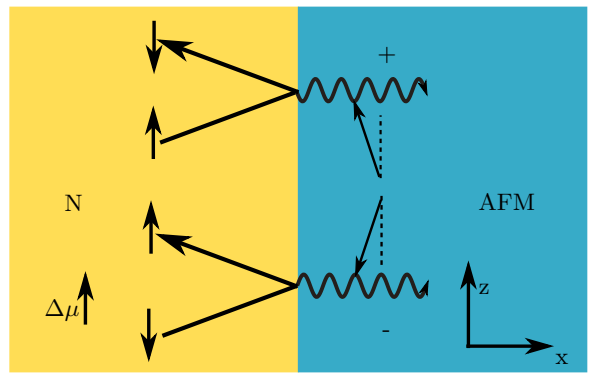
	\caption{Schematic representation of the system and the interactions that we study.
	An insulating antiferromagnet (right) is coupled to a normal metal conductor (left).
	The interface is parallel to the $yz$-plane.
	The antiferromagnet has an easy axis anisotropy along the $z$ axis.
	The positive $y$ direction points into the plane of the figure.
	Two types of magnons exist, which are related by a reflection in the $xy$-plane.}
	\label{fig:intro.system}
\end{figure}
The antiferromagnet is shaped as a rectangular cuboid aligned with the simple cubic lattice, and it contains $N$ sites. 
There is one spin at each site.
The axes of the lattice are aligned with those of the coordinate system defined in Fig.\ \ref{fig:intro.system}. 
We define two sublattices $A$ and $B$. 
Each sublattice contains half of the sites.
In the classical ground state, all spins in sublattice $A$ are aligned along $\hat{z}$, and all spins in sublattice $B$ point in the opposite direction. 
Both sublattices are face-centered cubic, and all nearest neighbors of a node in sublattice $A$ reside in sublattice $B$ and vice versa. 
We include spin-spin interactions between nearest neighbor atoms. 
Furthermore, we assume that the interaction between any pair of nearest neighbors is the same.

Including an easy-axis anisotropy along the $z$ axis, the antiferromagnet Hamiltonian becomes 
\begin{equation}
		H_{\Ins} = J \sum_{\nearnb{i}{j} \mid i,j \in \Ins} \vec{S}_i \cdot \vec{S}_j - K_z \sum_{i} S_{i z}^2 .
	\label{eq:dynamics.magnetHamiltonian}
\end{equation}
where $\nearnb{i}{j}$ is a pair of nearest neighbor sites.
The indices $i$ and $j$ uniquely identify lattice sites,
and $i \in \Ins$ indicates that site $i$ is in the antiferromagnet.

We refer to the $yz$-plane, where there is translation symmetry, as the transverse plane. 
Spin transport takes place along the $x$ direction.
The AFI has left and right boundaries: the left one is at the normal metal-AFI interface, and the right one is at the antiferromagnet-vacuum interface.
The distance between these boundaries, $N_x \A$, is the length of the AFI.
Here, $\A$ is the lattice constant.
We assume that the AFI is longer than the spin coherence length such that the magnon properties at the left and right boundaries of the AFI are independent. 
Similarly, the extent of the AFI in the transverse directions is assumed to be sufficiently large such that we can disregard the details of the transverse boundaries. 
The $y$ and $z$ directions have widths $N_y \A$ and $N_z \A$, respectively. 
We use periodic boundary conditions in the transverse plane. In this model, the two sublattices have the same reciprocal lattice and Brillouin zone. 
The magnon momenta are defined within this Brillouin zone.
The Brillouin zone of the full lattice is twice as large and contains twice as many possible momenta.

Using the Holstein-Primakoff transformation, we define two sets of magnon creation operators $a^{\dag}_i$ and $b^{\dag}_i$. 
Each operator corresponds to the excitation of a single spin at site $i$; $a$ and $b$ denote the corresponding sublattices. 
We expand the Hamiltonian in powers of the magnon operators and disregard terms involving four or more magnons. 
The Hamiltonian of  Eq.~(\ref{eq:dynamics.magnetHamiltonian}) only results in terms involving an even number of magnons.

We derive the energy eigenstates in Appendix \ref{sec:scattering.magnons}.  
Most of the energy eigenstates are delocalized and extend throughout the magnet. 
Additionally, there are surface states localized toward the interface.
The decay length of the surface states, $\lambda_{E_{\vec{q} }}$, is introduced in Appendix \ref{sec:scattering.magnons}.

Let us first consider surface states with energies, $E_{\vec{q} }$, that are of the same order as the gap, $E_0$. 
In this case, the decay length, $\lambda_{E_{\vec{q} }}$, is of the same order as the decay length of the magnon ground state $\lambda_{E_{0}}$. 
The latter is of the order $\A J/K_z$.
We assume that the anisotropy energy, $K_z$, is sufficiently small compared to the exchange energy, $J$, that $\A J/K_z$ is considerably longer than the length of the AFI, $N_x \A$.
For the antiferromagnetic material RbMn$\text{F}_{3}$, $\A J/K_z$ is on the order of $1$ mm\cite{windsor1966spin,cheng2015terahertz,eremenko1968anisotropy}.
When $\A J/K_z$ is considerably longer than $N_x \A$, the magnon surface states with energies on the order of the gap are approximately uniform throughout the AFI.
In Appendix \ref{sec:scattering.magnons}, we show that the uniform surface states are approximately equal to delocalized states with zero longitudinal momentum, $q_x = 0$. 
Consequently, we model the behavior of the surface states using the uniform delocalized states.

When the magnon energy, $E_{\vec{q} }$, is substantially higher than the gap, the surface state decay length, $\lambda_{E_{\vec{q} }}$, is of the same order as the length of the AFI.
In Appendices \ref{sec:scattering.magnons} and \ref{sec:scattering.coefficients}, we show that the magnon-electron interaction at the interface involving a surface state is comparable to that involving a delocalized state.
For the surface states, the longitudinal momentum $q_x$ is determined by the transverse momentum, $\vec{q}_{\perp} = (q_y,q_z)$, via the boundary conditions, whereas $q_x$ is a free parameter for the delocalized states. 
There are therefore many more delocalized states than surface states. 
The latter non-uniform surface states only constitute a small fraction of the high-energy magnons and will be disregarded.

With this information, we can model the behavior of all energy eigenstates using delocalized states, where the surface states are modeled as uniform delocalized states.
We define two sets of delocalized states in terms of the annihilation operators $\alpha_{\vec{q}}^{+}$ and $\alpha_{\vec{q}}^{-}$, respectively.
Here, $\vec{q}$ is the wavevector of the incoming part of the delocalized state, and the superscripts $+$ and $-$ refer to the type of magnon.
The $+$ ($-$) magnons carry a spin angular momentum of $\hbar$
pointing in the $-\hat z$ ($+\hat z$) direction.

The amplitude of a magnon in a delocalized state $\alpha_{\vec{q}}^{\pm}$ at node $i$ is described in terms of a wave function.
We express the wave function using two continuous functions in real space; each function provides the amplitudes associated with one sublattice.
By interchanging those functions in the state $\alpha^{+}_{\vec{q}}$ between the two sublattices, we obtain the wave function associated with the state $\alpha^{-}_{\vec{q}}$.
The interchange corresponds to a reflection of all the spins of our system in the $xy$-plane while simultaneously shifting the lattice by one lattice spacing in the transverse plane.
Because $H_{\Ins}$ of Eq.~(\ref{eq:dynamics.magnetHamiltonian}) is symmetric both under the reflection and under the shift, the two types of magnons have the same dispersion relation.
Therefore, the magnon ground state is degenerate; one degenerate state is of type $+$, and the other is of type $-$.

We focus on the low-energy, long-wavelength spin-wave excitations, $q \A \ll 1$.
The magnon dispersion for long wavelengths is derived in Appendix \ref{sec:scattering.magnons}.
The result is well known, 
\begin{equation}
	E_{\vec{q}} = \sqrt{(\hbar v q)^2+E_{0}^2} ,
	\label{eq:dynamics.dispersion}
\end{equation}
where the (high-energy) spin-wave velocity is $v= \sqrt{3} \hbar J \A$ and the spin-wave gap is\cite{hulthen1936uber}
\begin{equation}
	E_{0}=\hbar^2 \sqrt{6 J K_z} .
	\label{eq:dynamics.magnongap}
\end{equation}

\subsection{Properties of the normal metal}
We consider a normal metal with the same cubic lattice structure as the antiferromagnet. 
Deviations from such an ideal system cause a renormalization of the electron-magnon scattering rates but do not change the main physics; thus, they are not considered further.

In a similar way as for the AFI, we disregard the details of the boundaries in the transverse plane and use periodic boundary conditions. 
We assume that the scattering processes at the left and right boundaries of the normal metal are independent due to inelastic scattering in the bulk of the metal. 
We use a tight-binding model for the conduction electrons:
\begin{equation}
		H_{\Cond} = \sum_{\nearnb{i}{j} \mid i,j \in \Cond ,\sigma} -t \left( c^{\dag}_{\sigma i} c_{\sigma j} + c^{\dag}_{\sigma j} c_{\sigma i} \right) + 6t
	\label{eq:dynamics.metalHamiltonian}
\end{equation}
where $c_{\sigma i}^{(\dag)}$ annihilates (creates) an electron with spin $\sigma$ at node $i$.
Here, $i \in \Cond$ indicates that site $i$ is in the normal metal.
We assume half filling, and the Fermi energy is then $E_{\FE} = 6t$. 

\subsection{Antiferromagnet-metal coupling}
We consider a local exchange interaction, with energy scale $\hbar^2 J_{\Int}$, between the spins in the antiferromagnetic insulator and the itinerant spins in the normal metal along the interface:
\begin{equation}
		H_{\Int} = J_{\Int} \sum_{\nearnb{i}{j} \mid i \in \Ins, j \in \Cond} \vec{S}_i \cdot c^\dagger_{\sigma j}\boldsymbol\tau_{\sigma\sigma'}c_{\sigma' j} ,
	\label{eq:dynamics.InteractionHamiltonian}
\end{equation}
where $\vec{\tau}=(\tau_x,\tau_y,\tau_z)$ is a vector of Pauli matrices.
When the antiferromagnet is in its classical ground state, the exchange interaction of Eq.~(\ref{eq:dynamics.InteractionHamiltonian}) induces a static spin-dependent potential, $H_{\Int}^{(0)}$, seen by the itinerant electrons at the interface.
We determine the energy eigenstates of the electrons in the normal metal, including the interface potential $H_{\Int}^{(0)}$.
In Appendix \ref{sec:scattering.conduction}, we express the eigenstates as scattering states in a similar way as for the magnons in the antiferromagnet detailed in Appendix \ref{sec:scattering.magnons}. 

The annihilation operators of the scattering states are denoted as $c_{\sigma \vec{k}}$, where $\sigma$ is the spin $z$ component of the electron spin and $\vec{k}$ is a wavevector that is incoming to the AFI-N interface.
The conduction electron momenta are assumed to be near the Fermi surface, and they are defined within the Brillouin zone of the normal metal lattice. 

By expanding the interface exchange interaction of Eq.~(\ref{eq:dynamics.InteractionHamiltonian}) to the first order in the magnon operators, we find in Appendix \ref{sec:scattering.coefficients} an interaction of the form
\begin{equation}		
		H_{\Int} - H_{\Int}^{(0)} = \sum_{\vec{q},\vec{k}, \vec{k}'}  \left( V^{+}_{\vec{q},\vec{k}, \vec{k}'}  \alpha^{+}_{\vec{q}} + V^{-}_{\vec{q}, \vec{k}, \vec{k}'}  \alpha^{- \dag}_{\vec{q}} \right) c^{\dag}_{\spindown \vec{k}} c_{\spinup \vec{k}'} + \text{h.c.}
	\label{eq:dynamics.HamiltonianInterface}
\end{equation}
The sum over magnon and conduction electron momenta includes only the incoming wavevectors with respect to the interface.
We separate the wavevectors $\vec{k}$ into a part that is parallel to the direction of transport, $k_{\parallel} = k_x$, and a transverse part, $\vec{k}_{\perp} = (k_y,k_z)$.
We assume that the exchange energy, $J \hbar^2$, and the Fermi energy are large compared to the other energy scales of the problem.
In Appendix \ref{sec:scattering.coefficients}, we calculate the coefficients $V^{\pm}_{\vec{q},\vec{k}, \vec{k}'}$ for the modes that are uniform in the direction of transport and the modes with finite longitudinal momenta $q_{\parallel}$ separately, and we find that they differ by a factor\cite{kapelrud2013spin} of $\sqrt{2}$,
\begin{equation}		
		V^{\pm}_{\vec{q}, \vec{k}, \vec{k}' } = \begin{cases}
			U^{\pm}_{\vec{q}, \vec{k}, \vec{k}' } & q_{\parallel} \neq 0 \\
			\frac{1}{\sqrt{2}} U^{\pm}_{\vec{q}, \vec{k}, \vec{k}' } & q_{\parallel} = 0 
		\end{cases} \, .
	\label{eq:interface.VcoefficientFactor}
\end{equation}
The difference in the interface electron-magnon coupling of Eq.~(\ref{eq:interface.VcoefficientFactor}) is the reason why the enhanced Gilbert damping for the longitudinal finite wavelength modes is twice that of the longitudinal homogeneous mode\cite{kapelrud2013spin}.

In the limit of a large exchange energy and a large Fermi energy, we can show that the dominant contributions to the coefficients are
\begin{widetext}
\begin{equation}		
		U^{\pm}_{\vec{q}, \vec{k}, \vec{k}' } = \frac{ \sqrt{2} \hbar^2 J_{\Int}}{M_x \sqrt{N} } u_{\vec q} \frac{\sin\left(k'_x \A\right) \sin\left(k_x \A \right) }{\left(\lambda^2 + 1 \right)^2 } \Biggl( \delta^{\perp}_{ \vec{k}^{\prime \U} , \vec{k} \pm \vec q } \left( e^{ \pm i (k_x - k'_x) \A } - \lambda^2 \right) + \delta^{\perp}_{ \vec{k}' , \vec{k} \pm \vec q } \lambda \left( e^{ \pm i k_x \A } - e^{ \mp i k'_x \A } \right) \Biggr) ,
	\label{eq:interface.Vcoefficients}
\end{equation}
\end{widetext}
as shown in Appendix \ref{sec:scattering.coefficients}.
Here, $\delta^{\perp}_{\vec{p},\vec{q}}=\delta_{\vec{p}_{\perp},\vec{q}_{\perp}}$ are Kroenecker deltas for the transverse vector components, and $\lambda = \hbar^2 J_{\Int}/(4 t)$.
The parameter $u_{\vec{q}} \approx \left( 3 \hbar^2 J/ (2 E_{\vec{q} } ) \right)^{1/2}$ is introduced in Appendix \ref{sec:scattering.magnons} and is equal to the Bogoliubov coefficient of the bulk model.
Furthermore, we have defined the Umklapp-scattered momentum $\vec{k}^{\U}$ such that $\vec{k}^{\U}_{\perp} = \vec{k}_{\perp} + \pi / \A \left(\uvec{y} + \uvec{z} \right)$.
We define $k^{\U}_{\parallel}$ in terms of energy conservation, $\varepsilon_{\vec{k}} = \varepsilon_{\vec{k}^U}$.

The amplitudes, $V^{\pm}_{\vec{q}, \vec{k}, \vec{k}' }$, of Eq.~(\ref{eq:interface.Vcoefficients}) are directly proportional to the Bogoliubov coefficient $u_{\vec{q}}$.
Because the exchange energy is considerably larger than the relevant magnon energies, $u_{\vec{q}} \gg 1$. 
Thus, the inclusion of this factor strongly enhances the total transport rates. 
The amplitudes $V^{\pm}$ describe processes where a conduction electron creates or annihilates a magnon.
The delta functions in the expression for $V$ ensure that the momentum of the conduction electrons is conserved in the transverse plane, but they also allow for a shift by a large momentum $\pi/\A \left(\uvec{y} + \uvec{z} \right)$. 
We refer to the processes where transverse momentum is conserved as normal scattering processes and the processes where the transverse momentum is shifted as Umklapp scattering processes.
The terms proportional to $\lambda$ in Eq.~(\ref{eq:interface.Vcoefficients}) are caused by the proximity effect.
When $\lambda \ll 1$, the proximity effect is negligible and the total rate of scattering involving magnons is low.
When $\lambda \ll 1$ and electron scattering with a magnon does occur, the Umklapp scattering process dominates the normal scattering process.

\section{\label{sec:rates}Transport rates}

We assume that the electron-magnon interaction at the interface is weak, and we treat it as a perturbation with respect to the decoupled systems of electrons and magnons.
We now calculate the rate of change of the number of magnons of each type, $I^{\pm}$, caused by the electron-magnon coupling at the interface.
Using Fermi's golden rule, we find that
\begin{align}		
		I^{\pm} &= \!\frac{2 \pi}{\hbar} \!\!\sum_{\vec{q}, \vec{k}, \vec{k}'} \!\!\left[ \Tr \left\{ \rho  V^{\pm}_{\vec{q}, \vec{k}, \vec{k}'}  \alpha^{\pm}_{\vec{q}} c^{\dag}_{\spindown \vec{k}} c_{\spinup \vec{k}'} V^{\pm *}_{\vec{q}, \vec{k}, \vec{k}'} (\alpha^\pm_{\vec{q}})^\dag  c^{\dag}_{\spinup \vec{k}'} c_{\spindown \vec{k}}  \right\} \right.\notag \\
		- &\left.\!\!\Tr \left\{\! \rho V^{\pm*}_{\vec{q}, \vec{k}, \vec{k}'} (\alpha^\pm_{\vec{q}})^\dag  c^{\dag}_{\spinup \vec{k}'} c_{\spindown \vec{k}} V^{\pm}_{\vec{q}, \vec{k}, \vec{k}'} \alpha^{\pm}_{\vec{q}} c^{\dag}_{\spindown \vec{k}} c_{\spinup \vec{k}'} \!\right\}\! \right] \!\delta \!\left( E_{ \Fin } {-} E_{ \Ini } \right) \!.
	\label{eq:rates.FermiCurrent}
\end{align}
In Eq.~(\ref{eq:rates.FermiCurrent}), $\rho$ is the density matrix of the decoupled systems of electrons and magnons. 
The trace involves a sum over all quantum states. 
The first line of Eq.~(\ref{eq:rates.FermiCurrent}) represents the creation of magnons from an initial state, and the second line describes the annihilation of magnons.
The difference between the creation and annihilation rates determines the magnon number rate of change $I^{\pm}$. 
The creation and annihilation of magnons causes a change in the spin angular momentum of the itinerant electrons.
$E_{ \Fin }$ and $E_{\Ini}$ are the final and initial state energies of the non-interacting model, respectively.

The magnons and conduction electrons are also affected by other interactions in the bulk of the materials. 
These interactions may be between the magnons or between the electrons, but there could also be interactions with other degrees of freedom, such as phonons. 
We consider bulk interactions that conserve the number of magnons and electrons. 
In other words, we assume that the magnon relaxation time is sufficiently long such that the magnon distributions can be experimentally observed.
We use the phrase magnon number conserving process to describe interactions where the numbers of magnons of type $+$ and $-$ are conserved separately. 
Treating the electron-magnon interaction at the interface as a perturbation, we assume that the bulk interactions are considerably faster than the interface interactions. 
Furthermore, we assume that any bulk interactions that create or destroy magnons are slower than the interface interactions. 
The assumption that the magnon number conserving interactions are dominant is valid for the magnons in the ferrimagnet yttrium-iron-garnet (YIG)\cite{cornelissen2016magnon}.

The symmetry of the Hamiltonian of Eq.~(\ref{eq:dynamics.magnetHamiltonian}) under rotations in spin space around the $z$ axis implies conservation of the total spin angular momentum along the $z$ direction.
When the total spin angular momentum of the magnons is conserved, interactions conserving the total number of magnons also conserve the number of magnons of types $+$ and $-$ separately. 
The terms in the Hamiltonian of Eq.~(\ref{eq:dynamics.magnetHamiltonian}), where the magnon number is not conserved, simultaneously create (annihilate) both a $+$ and a $-$ magnon.
This does not violate the conservation of angular momentum since the combined spin angular momenta of the two magnons in such pairs vanish.
We disregard the processes where pairs of magnons are created or annihilated from our model.
The remaining magnon-magnon interactions are examples of magnon number preserving bulk interactions.
Similarly, we assume that the magnon number is conserved in the magnon-phonon interactions.

The bulk interactions can drive the magnons into a quasi-equilibrated distribution in a normal phase or a condensate phase. 
For both the normal phase and the condensate phase, there are two thermal baths of magnons  with a Bose-Einstein distribution: one for each type of magnon.

The electrons are assumed to be quasi-equilibrated with a Fermi-Dirac distribution, but we allow for a difference in the chemical potentials $\mu_{\spinup}$ and $\mu_{\spindown}$ of the spin-up and spin-down electrons. 
We define the spin accumulation as $\Delta \mu = \mu_{\spinup} - \mu_{\spindown}$. 
The temperature of the conduction electrons is assumed to be independent of the spin.
We will show that the two types of magnons are affected differently by the spin accumulation.

The interactions across the interface are slow compared to the decoherence time of each subsystem. 
Therefore, the density matrix of the coupled system is well approximated by a decoupled density matrix of the form $\rho = \rho_{\Ins} \otimes \rho_{\Cond}$.
Here, $\rho_{\Ins}$ is the density matrix of the antiferromagnet, and $\rho_{\Cond}$ is the density matrix of the normal metal. 
The quasi-equilibrated thermal states of the magnons and conduction electrons are described by the density matrices $\rho_{\Ins}$ and $\rho_{\Cond}$, where
\begin{align}
	 \Tr \left\{ \rho_{\Ins} \left( \alpha^{\pm}_{\vec{q}} \right)^{\dag} \alpha^{\pm}_{\vec{q}'}\right\} = & \delta_{\vec{q}, \vec{0}} \delta_{\vec{q}', \vec{0}} n^{\pm}_{0} \notag \\
	       & + \BE \left( \beta_{\pm} \left(E_{\vec{q}} - \mu_{\pm} \right) \right) \delta_{\vec{q}, \vec{q}'}, \label{eq:rates.thermalStates} \\
		\Tr \left\{ \rho_{\Cond}  c^{\dag}_{\sigma \vec{k}} c_{\sigma' \vec{k}'}\right\} = & \FD \left( \beta_{\Cond} \left(\varepsilon_{\vec{k}} - \mu_{\sigma} \right) \right) \delta_{\vec{k}, \vec{k}'} \delta_{\sigma, \sigma'}. \notag
\end{align}
Here, $\sigma$ is either spin up ($\spinup$) or spin down ($\spindown$), and $\varepsilon_{\vec{k}}$ is the conduction electron energy.
The presence of a Bose-Einstein condensate implies a macroscopic number of magnons in one or both of the magnon ground states, $n_0^\pm$. 
In Eq.~(\ref{eq:rates.thermalStates}), $\mu_{+}$ ($\mu_-$) is the chemical potential of the magnons of type $+$ ($-$), and $\beta_{\pm}=1/\left(\kB T_{\pm}\right)$, where $T_{+}$ ($T_{-}$) is the effective temperature of the magnons of type $+$ ($-$). 
Finally, $\beta_{\Cond}=1/\left(\kB T_{\Cond}\right)$, where $T_{\Cond}$ is the temperature of the conduction electrons.

We group the magnons into four contributions: the thermal magnons of types $+$ and $-$, and the two possible condensates.
The number of magnons in each group is denoted by $n^{\pm}_{\Th}$ and $n^{\pm}_{0}$, respectively.

With this information, we will now compute the rate of change of the magnon numbers $n^{\pm}_{\Th}$ due to electron-magnon scattering. 
To this end, we use Eq.~(\ref{eq:rates.FermiCurrent}) for the magnon current. 
We assume that the Fermi energy of the conduction electrons and the exchange energy of the antiferromagnet are both considerably larger than the other relevant energy scales, such as the spin accumulation and the magnon gap. 
The dominant contribution to $I^{\pm}_{\Th}$ is 
\begin{align}		
I^{\pm}_{\Th} &= \frac{\pi}{\hbar} \int g^{\pm}_{\Ins} \left( E \right) g_{\Cond}^2 \abs{u_{E}}^2 V^{\pm}\left( E \right) \left( E \pm \Delta \mu \right) \notag \\
&\phantom{{}=1} \left( \BE \left( \beta_{\Cond} \left(E \pm \Delta \mu \right) \right) - \BE \left( \beta_{\pm} \left(E - \mu_{\pm} \right) \right) \right) \diff{E} .
	\label{eq:rates.ThermalCurrent}
\end{align}
In this expression, the coefficient $u_E$ that appears is defined in the following way. 
We define $q_E$ as the length of a wavevector $\vec{q}$, where $E_{\vec{q}} = E$ and $u_{E}$ is the value of $u_{\vec{q}}$ when $\abs{\vec{q}} = q_E$. 
To the leading order, the Bogoliubov parameter $u_{\vec{q}}$ and the magnon energy $E_{\vec{q}}$ only depend on $\vec{q}$ through its magnitude.

In Eq.~(\ref{eq:rates.ThermalCurrent}), we have introduced 
\begin{align}		
V^{\pm} \left( E \right) &= \frac{V_{\Cond}^2 V_{\Ins}}{\left( 2 \pi\right)^9 g^{\pm}_{\Ins} \left( E \right) g_{\Cond}^2} \iiint \abs{V^{\pm}_{\vec{k}, \vec{k}', \vec{q}}}^2 \frac{\delta \left( E - E_{\vec{q}}\right)}{\abs{u_{E}}^2} \notag \\
&\phantom{{}=1}    \delta\left( \varepsilon_{\vec{k}} - E_{\FE}\right)  \delta\left( \varepsilon_{\vec{k}'} - E_{\FE} + E \right)  \diffv{\vec{k}} \diffv{\vec{k}'}  \diffv{\vec{q}} .
	\label{eq:rates.VEnergy}
\end{align}
Here, $V_{\Ins} = N \A^3$ is the volume of the antiferromagnet and $V_{\Cond} = M \A^3$ is the volume of the normal metal,  $g^{+}_{\Ins}$ $(g^{-}_{\Ins})$ is the density of states of the magnons of type $+$ $(-)$, and $g_{\Cond}$ is the conduction electron density of states at the Fermi surface. 

Similarly, the current into the condensate is
\begin{equation}		
I^{\pm}_0 = \frac{\pi}{\hbar} g_{\Cond}^2 \abs{u_{\vec{0}}}^2 V^{\pm}_0 \left( - E_{0} \mp \Delta \mu \right) n^{\pm}_0,
	\label{eq:rates.CondensateCurrent}
\end{equation}
where 
\begin{align}		
V^{\pm}_0 &= \frac{V_{\Cond}^2 }{\left( 2 \pi\right)^6 g_{\Cond}^2} \frac{2}{\abs{u_{\vec{0}}}^2} \iint  \abs{V^{\pm}_{\vec{k}, \vec{k}', \vec{0}}}^2 \notag \\ 
&\phantom{{}=1} \delta\left( \varepsilon_{\vec{k}} - E_{\FE}\right) \delta\left( \varepsilon_{\vec{k}'} - E_{\FE} + E_{0} \right) \diffv{\vec{k}} \diffv{\vec{k}'} .
	\label{eq:rates.def.Vground}
\end{align}
The energy integral in Eq.~(\ref{eq:rates.ThermalCurrent}) runs over the energy of the magnons that are created or annihilated at the interface.
Only magnons with energies that are considerably smaller than the exchange energy and the Fermi energy contribute to the current $I^{\pm}_{\Th}$.
Therefore, it is sufficient to evaluate $V^{\pm} \left( E \right)$ in the limit where $E$ is substantially smaller than the exchange energy and the Fermi energy.
In this limit, we find that
\begin{align}	
V^{\pm} \left( E \right) &\approx 2 V^{\pm}_0 \approx \frac{ \pi^2 N_{\perp} J^{2}_{\Int} \hbar^4 }{ M^2 N } \frac{ \left(\lambda^2 + 1 \right)^2 0.12 + \lambda^2 0.40 }{\left(\lambda^2 + 1 \right)^4 },
	\label{eq:rates.Vapproximate}
\end{align}
where $N_{\perp}$ is the number of lattice nodes in one transverse layer.

Note that the currents of Eq.~(\ref{eq:rates.CondensateCurrent}) and Eq.~(\ref{eq:rates.ThermalCurrent}) are proportional to the large factor $\abs{u_{E}}^2 = 3 \hbar^2 J/(2 E)$. 
These factors do not occur in the corresponding expressions for a ferromagnetic system\cite{PhysRevLett.108.246601}. 
The implication is that the electron-magnon coupling at the interface is considerably stronger in antiferromagnets than in ferromagnets.
Consistent with this finding, a similar enhancement occurs in the heat transfer across AFI-N interfaces\cite{brataas2015heat}.

The density of states for the conduction electrons is approximately constant because the dynamics involve only electrons near the Fermi surface.
The magnon density of states, however, vanishes near the ground state energy.
The densities of states of the conduction electrons and the magnons are respectively
\begin{align}		
g_{\Cond} \approx \frac{M}{2 t \left(2 \pi\right)^3 } 17.695 , \quad g^{\pm}_{\Ins} \left( E \right) \approx \frac{V_{\Ins}}{4 \pi^2} \frac{E \abs{q_E}}{\left( \hbar v \right)^2}.
	\label{eq:rates.dos}
\end{align}
We assume periodic boundary conditions when estimating the densities of states of the AFI.

In Appendix \ref{sec:scattering.magnons}, we introduce a reflection angle, $\phi_{\vec{q}}$, that determines the magnon amplitude near the interface.
We find that the reflection angle is much smaller than $1$ for all delocalized states with energy $E$ on the order of the gap.
For magnons with energies that are much larger than the gap, we can disregard the gap in the dispersion relation, Eq.~(\ref{eq:scattering.magnons.dispersion}).
Using the approximate dispersion, we find that the reflection angle is of the order $q^2 \A^2 / \abs{q_x} \A $. 
Typically, since the magnon wavelength is considerably longer than the lattice spacing, the reflection angle is also much smaller than $1$ in this regime. 
However, there are exceptions when the momentum vector $\vec{q}$ is almost parallel to the interface, e.g., when $q_x$ is relatively small. 
Nevertheless, only a small portion of the delocalized states in the thermal bath  have momenta parallel to the interface since the magnon dispersion is isotropic. 
Because the delocalized states with finite reflection angles only constitute a small portion of the thermal cloud, and interact slower with the conduction electrons than the rest of the thermal bath, their contributions to the magnon currents in Eq.~(\ref{eq:rates.VEnergy}) are negligible.

\section{\label{sec:fixedT}Condensate and Instability}

The two types of magnons can both condense, and they also form separate thermal clouds, which could have different distributions.
We assume that the temperatures of both clouds remain fixed by an external reservoir, such as phonons.
We denote the common magnon temperature as $T_{\Ins}$.

First, we consider the steady-state {\it normal phase} where no condensates are present. 
In this regime, the magnon chemical potentials for the two types together with the temperatures determine the magnon distribution. 
The magnon chemical potentials are smaller than the magnon gap, $\mu_{\pm} < E_0$. 
The vanishing of the net currents of magnons into the antiferromagnet of Eq.~(\ref{eq:rates.ThermalCurrent}) determines the chemical potentials. 
We directly observe from Eq.~(\ref{eq:rates.ThermalCurrent}) that when there is no thermal bias, $T_{\Cond} = T_{\Ins}$, the magnitudes of the chemical potentials equal the spin accumulation, $\mu_{\pm} = \mp \Delta \mu$. 

A finite thermal bias causes the magnitudes of the chemical potentials to deviate from the spin accumulation.
We consider the case where $T_{\Ins}$ is large compared to the magnon gap $E_0$ and the relative difference between $T_{\Cond}$ and $T_{\Ins}$ is small. 
In this limit, we find that the chemical potentials are 
\begin{equation}		
	\mu_{\pm} = \mp \Delta \mu + \frac{ 18 \zeta \left( 3 \right)}{\pi^2} \kB \left(T_{\Cond} - T_{\Ins} \right) \, .
	\label{eq:fixedT.chemicalPotential}
\end{equation}
The Riemann zeta function is $\zeta \left( n \right)=\sum_{k=1}^{\infty} 1 / k^{n} $ for integral numbers $n$.

When the magnon chemical potential for one type approaches the magnon gap, the number of magnons in the associated ground state becomes vast and macroscopic, creating a condensate. 
However, the chemical potential never increases beyond the magnon gap. 
In the {\it condensate phase}, one of the magnon chemical potentials, $\mu_{+}$ or $\mu_{-}$, is equal to the gap, $E_0$. 
In the condensate phase, we also need to determine the number of condensate magnons, $n_0^+$ and $n_0^-$.

A spin accumulation in the normal metal increases the number of magnons of one type and decreases the number of magnons of the other type. 
This asymmetry causes the creation of only one condensate.  
Without a loss of generality, we assume that the spin accumulation $\Delta \mu$ is negative, which implies that only the condensate of type + is present. 
The dynamics of the system can now be described by
\begin{subequations}
\begin{align}		
	\dtime{n}^{+}_{\Th} &= I^{+}_{\Th} + I^{+}_{\Sec}, \\
	\dtime{n}^{-}_{\Th} &= I^{-}_{\Th},  \\
	\dtime{n}^{+}_{0} &= I^{+}_0  - I^{+}_{\Sec} ,
\end{align}
\label{eq:fixedT.numbers}
\end{subequations}
where $I^{+}_{\Sec}$ is the net magnon transfer rate from the condensate into the thermal cloud. 
$I^{+}_{\Sec}$ is caused by magnon number conserving bulk interactions. 
Next, we will find steady-state solutions of the dynamical equations (\ref{eq:fixedT.numbers}) and the threshold values of the spin accumulation and temperature for inducing a condensate.

When the spin accumulation is larger than the gap, swasing can occur. 
The electron-magnon coupling-induced current of Eq.~(\ref{eq:rates.CondensateCurrent}) is proportional to the number of magnons already present in the condensate.
This feature can lead to an exponential increase in the number of magnons in the condensate.
It is this phenomenon that is a swasing instability\cite{berger1996emission,PhysRevLett.108.246601,bender2014dynamic}. 
The direction of the current of Eq.~(\ref{eq:rates.CondensateCurrent}) changes when the spin accumulation increases beyond the gap.
Beyond the swasing instability, when the spin accumulation is larger than the gap, the number of magnons in the condensate rapidly increases, forming a large condensate. 
The growth of the condensate will continue until higher-order interactions between the magnons prevent a further population build-up.

However, a large spin accumulation $- \Delta \mu > E_0$ does not necessarily lead to swasing. 
To study this further, we need to take into account interactions between the thermal bath of magnons and the condensate, given by $I^{+}_{\Sec}$.
Such bulk interactions are fast and ensure that the magnons in the cloud remain in a non-equilibrium thermal distribution.
Even when the spin accumulation is larger than the gap, there could be stable steady-state solutions where $n_{0}^{+} = 0$.
When $n_{0}^{+} = 0$, the system is in the normal phase.
The normal phase is stable when the thermal cloud current of Eq.~(\ref{eq:rates.ThermalCurrent}) vanishes for a magnon chemical potential that is smaller than the gap, $\mu^{+} < E_{0}$.
This condition for the stability of the normal phase state does not depend on the relative magnitude of the spin accumulation and the gap.
Therefore, Eq.~(\ref{eq:fixedT.chemicalPotential}) is valid even for large spin accumulations as long as both $\mu_{+}$ and $\mu_{-}$ are below the threshold, $E_0$, when we maintain the assumption that $T_{\Ins}$ is much larger than $E_0 / \kB$ and $\abs{T_{\Ins} - T_{\Cond}}$.
As shown in Eq.~(\ref{eq:fixedT.chemicalPotential}), the normal phase remains stable even when the spin accumulation is larger than the gap provided that the conduction electron temperature $T_{\Cond}$ is smaller than the magnon temperature $T_{\Ins}$.
However, when $T_{\Cond}$ is larger than $T_{\Ins}$ and $- \Delta \mu > E_0$, the normal phase is unstable and swasing will occur. 
The cases when $- \Delta \mu > E_0$ and a normal phase state is a possible steady state solution are hysteretic\cite{bender2014dynamic}. 
Whether swasing or a normal phase occurs depends on the past history. 

We now further consider the case in which the spin accumulation is smaller than the gap, $- \Delta \mu < E_0$.
If the conduction electron temperature $T_{\Cond}$ is larger than that of the magnons, $T_{\Ins}$, the normal phase may become unstable even though the spin accumulation is small.
In other words,  a condensate can form for spin accumulations that are too small to induce swasing.
When no stable normal phase state exists, the current $I^{+}_{\Th}$ of Eq.~(\ref{eq:rates.ThermalCurrent}) is always positive, and the chemical potential $\mu^{+}$ will increase until it approaches $\mu^{+} = E_0$.
When $\mu^{+} = E_0$, $I^{+}_{\Th}$ can be fully determined using Eq.~(\ref{eq:rates.ThermalCurrent}). 
We will denote this current by $I_{\Th 0}$. 
In the limit where $T_{\Ins}$ is considerably larger than $E_0 / \kB$ and $\abs{T_{\Ins} - T_{\Cond}}$, we find that
\begin{align}
	I_{\Th 0} &= \frac{\pi}{\hbar} V^{+} \left( E_0 \right) g_{\Cond}^2 \frac{V_{\Ins}}{8 \sqrt{3} \pi^2} \frac{\kB^2 T_{\Ins}^2}{\hbar^2 v^2} \notag \\
	&\left( \pi^2 \left(- E_0 - \Delta \mu \right) + 18 \zeta \left( 3\right)  \kB \left(T_{\Cond} - T_{\Ins} \right)\right).
	\label{eq:fixedT.condensateThermalCurrent}
\end{align}
Furthermore, the number of thermal magnons is fixed. 
In this case, using the conservation of the number of thermal magnons $n_Q^+$, Eq.~(\ref{eq:fixedT.numbers}) and Eq.~(\ref{eq:rates.CondensateCurrent}), we find that
\begin{equation}		
	\dtime{n}^{+}_{0} = \frac{\pi}{\hbar} g_{\Cond}^2 V^{+}_0 \abs{u_{\vec{0}}}^2 \left( - E_{0} - \Delta \mu \right) n^{+}_0 + I_{\Th 0}.
	\label{eq:fixedT.condensateEqOfMotion}
\end{equation}
We define a time scale $\frac{1}{\tau_{0}} = \frac{\pi}{\hbar} g_{\Cond}^2 V^{+}_0 \abs{u_{\vec{0}}}^2 \left( E_{0} - \Delta \mu \right)$, and we use this time scale to express the solution of Eq.~(\ref{eq:fixedT.condensateEqOfMotion}),
\begin{equation}		
	n^{+}_{0} \left( t \right) = \tau_{0}I_{\Th 0} + \left(n^{+}_{0} \left( 0 \right) - \tau_{0}I_{\Th 0} \right) e^{-\frac{t}{\tau_{0}}} .
	\label{eq:fixedT.condensateSolution}
\end{equation}
From Eq.\ (\ref{eq:fixedT.condensateSolution}), we observe that the steady-state number of condensed magnons is $n^{+}_{0} = \tau_{0}I_{\Th 0}$ and independent of the initial conditions. 
In this steady state, $I^{+}_0$ is negative, which indicates that magnons are leaving the condensate across the interface. 
However, the condensate is maintained by the transfer from the thermal cloud. 
This is caused by the bulk interactions that redistribute the magnons from the cloud and into the condensate. 
In turn, the spin accumulation supplies magnons to the thermal cloud. 
This steady state is a magnon Bose-Einstein condensate\cite{bender2014dynamic}.

\section{\label{sec:conclusions} Conclusions}

We have explored the prospect of Bose-Einstein condensation of magnons in insulating antiferromagnets coupled to normal metals.
Condensation occurs when large numbers of magnons are created at the interface.
The creation of magnons at the interface can be stimulated by a spin accumulation in the normal metal and a temperature difference across the interface.
Starting from a quantum-mechanical model, we describe the dynamics of the antiferromagnet in terms of two types of magnons, which carry the same energy and opposite spin angular momentum.
Both types of magnons strongly interact with the conduction electrons in an adjacent normal metal compared to magnons in ferromagnets.

The spin accumulation, which can be induced via the spin Hall effect, causes an imbalance of the magnon distribution between the two types of magnons.
For large imbalances, a coherent magnon condensate is created in one of the two degenerate uniform magnon ground states.
In contrast to the ferromagnetic model, a condensate can form for both signs of the spin accumulation.

\begin{acknowledgments}
	The research leading to these results has received funding from the European Research Council via Advanced Grant number 669442 ``Insulatronics''. 
\end{acknowledgments}

\appendix

\section{Magnon Eigenstates}
\label{sec:scattering.magnons}

We will now calculate the magnon eigenstates of the semi-infinite Heisenberg antiferromagnet. 
In a similar way, Appendix \ref{sec:scattering.conduction} presents the computation of the electron eigenstates in the normal metal. 
Finally, we use the magnon and electron eigenstates to compute the interface electron-magnon coupling in Appendix \ref{sec:scattering.coefficients}.
We are only interested in magnons with energies that are considerably lower than the exchange energy. 
However, to ensure that we find the correct solutions for small but finite magnon energies, we first find all magnon eigenstates and then take the low-energy limit.

We define the sublattices $A$ and $B$ as described in Sec.~\ref{sec:dynamics}.
We represent the spin operators at each lattice site in terms of boson operators by using the Holstein-Primakoff transformation,
\begin{subequations}
	\begin{align}
			S_{i x} + i S_{i y} &= \begin{cases}
													\hbar \sqrt{1 - a^\dag_{i} a_{i} } a_{i}  & i \in A \\
													\hbar b^{\dag}_{i} \sqrt{1 - b^{\dag}_{i} b_{ i} } & i \in B
										\end{cases} \, , \\
			S_{i x} - i S_{i y} &= \begin{cases}
													\hbar a^{\dag}_{i} \sqrt{1 - a^\dag_{i} a_{i} } & i \in A \\
													\hbar \sqrt{1 - b^{\dag}_{i} b_{ i} } b_{ i} & i \in B
										\end{cases} \, ,  \\
			S_{i z} &= \begin{cases}
										\hbar \left( \frac{1}{2} - a^\dag_{i} a_{i}\right) & i \in A \\
										\hbar \left( - \frac{1}{2} + b^{\dag}_{ i} b_{ i}\right) & i \in B 
									\end{cases} \, ,
	\end{align} 
	\label{eq:scattering.magnons.HolsteinPrimakoff}%
\end{subequations}
where $a_i$ annihilates a magnon on site $i$, which belongs to sublattice $A$, and $b_i$ is the corresponding annihilation operator for a magnon at site $i$ in the $B$ sublattice.
We assume that all the localized spins in the antiferromagnet have quantum number $1/2$.
Other quantum numbers can be accounted for by redefining the exchange and anisotropy energies.
The index $i$ can be decomposed into three integer indices $i = (i_x, i_y, i_z)$ corresponding to the location of a lattice site along the three axes $x$, $y$ and $z$.
The sites where $i_x < 0$ are in the normal metal, and the sites in the AFI satisfy $i_x \geq 0$. 
In real space, the lattice site $i$ is at coordinate $\vec{r}_{i} = (x_i,y_i,z_i)$.
We choose the coordinate system such that $x_i = \xN + i_x \A$ in the AFI and $x_i = \xN + \A - \D + i_x \A$ in the normal metal, where $\D$ is the thickness of the interface.
The nodes in the antiferromagnet that are adjacent to the interface are located in the transverse plane $x = \xN$.
Similarly, nodes, $i$, in the normal metal that are adjacent to the interface satisfy $x_i = \xN - \D$.
Furthermore, we define transverse coordinates as $\vec{r}_{i \perp} = (y_i,z_i) = \A (i_y, i_z)$.
Analogously, for wavevectors $\vec{p}$, we define the transverse wavevector $\vec{p}_{i \perp} = (p_y,p_z)$.
We also choose the indices such that the node $i = (0,0,0)$ is on sublattice $A$.

To second order in the boson operators, the Hamiltonian of Eq.~(\ref{eq:dynamics.magnetHamiltonian}) becomes
\begin{align}
		H_{\Ins} &= \frac{J \hbar^2}{2} \sum_{\nearnb{i}{j} \mid i,j \in \Ins} \Biggl(a_{i} b_{j} +  a^{\dag}_{i} b^{\dag}_{j} - \frac{1}{2} + a^\dag_{i} a_{i} + b^{\dag}_{ j} b_{j} \Biggr) \notag \\
		&- K_z \hbar^2 \sum_{i \in A}  \left( \frac{1}{4} - a^\dag_{i} a_{i}\right) - K_z \hbar^2 \sum_{i \in B} \left( \frac{1}{4} - b^{\dag}_{ i} b_{i}\right) .
	\label{eq:scattering.magnons.Hamiltoniannode}
\end{align}
As discussed in Sec.~\ref{sec:dynamics}, we use periodic boundary conditions in the transverse plane. 
We assume that the longitudinal length is larger than the inelastic scattering length and disregard the boundary conditions at the right boundary opposite to the interface.
In this section, we will determine the energy eigenstates of the magnons. 
We find the magnon eigenstates with the ansatz
\begin{subequations}
	\begin{align}
			\alpha^{+}_{\vec{p}} &= \sum_{i \in A} \frac{ u_{\vec{p}}\left( x_i \right) }{\sqrt{N/2}} e^{-i \vec{p}_{\perp} \vec{r}_{i \perp}} a_{i} - \sum_{i \in B} \frac{v_{\vec{p}}\left( x_i \right)}{\sqrt{N/2}} e^{i  \vec{p}_{\perp} \vec{r}_{i \perp}} b^{\dag}_{i}, \\
			\alpha^{-}_{\vec{p}} &= \sum_{i \in B} \frac{ u_{\vec{p}}\left( x_i \right) }{\sqrt{N/2}} e^{- i \vec{p}_{\perp} \vec{r}_{i \perp}} b_{i} - \sum_{i \in A} \frac{v_{\vec{p}} \left( x_i \right) }{\sqrt{N/2}} e^{i \vec{p}_{\perp} \vec{r}_{i \perp} } a^{\dag}_{i}  \, .
	\end{align}
	\label{eq:scattering.magnons.def.creationannihilation}%
\end{subequations}
We have to determine the longitudinal ($x$) dependence of the energy eigenstates represented by the functions $u_{\vec{q}}\left( x_i \right)$ and $v_{\vec{q}}\left( x_i \right)$.

The Hamiltonian of Eq.~(\ref{eq:scattering.magnons.Hamiltoniannode}) relates the functions $u_{\vec{q}}\left( x \right)$ and $v_{\vec{q}}\left( x \right)$ by
\begin{subequations}
	\begin{align}
	0 &= \left[(3J + K_z)\hbar^2 - E_{\vec{q}}\right] u_{\vec{q}}(x_i) +  \\
	&\phantom{{}=1} \frac{ J \hbar^2}{2} \left( 2 \gamma^{\perp}_{\vec{q}_{\perp}} v_{\vec{q}}^{*}(x_i) + v_{\vec{q}}^{*}(x_i-\A) + v_{\vec{q}}^{*}(x_i+\A) \right), \notag \\
	0 &= \left[(3J + K_z)\hbar^2 + E_{\vec{q}}\right]	v_{\vec{q}}(x_i) + \\
	&\phantom{{}=1} \frac{J\hbar^2}{2} \left( 2 \gamma^{\perp}_{\vec{q}_{\perp}} u_{\vec{q}}^{*}(x_i) + u_{\vec{q}}^{*}(x_i-\A) + u_{\vec{q}}^{*}(x_i+\A) \right) \notag
\end{align}
	\label{eq:scattering.magnons.bulkEquation}%
\end{subequations}
for all nodes, $i$, in the bulk of the AFI. 
We have defined 
\begin{subequations}
	\begin{align}
		\gamma_{\vec{q}} = \cos \left(q_x \A \right)  + \cos \left(q_y \A \right) + \cos \left(q_z \A \right), \label{eq:scattering.magnons.def.gammabulk} \\
		\gamma^{\perp}_{ \vec{q}_{\perp}} = \cos \left(q_{y} \A \right) + \cos \left( q_{z} \A \right) \label{eq:scattering.magnons.def.gammaperp}
	\end{align}
	\label{eq:scattering.magnons.def.gamma}%
\end{subequations}
At the normal metal-AFI boundary, we find the conditions
\begin{subequations}
	\begin{align}
		0 &= \left[(\frac{5}{2} J + K_z)\hbar^2 - E_{\vec{q}}\right] u_{\vec{q}}(\xN) +  \\
		&\phantom{{}=1} \frac{ J \hbar^2}{2} \left( 2 \gamma^{\perp}_{ \vec{q}_{\perp}} v_{\vec{q}}^*(\xN) + v_{\vec{q}}^*(\xN+\A) \right), \notag \\
		0 &= \left[(\frac{5}{2}J + K_z)\hbar^2 + E_{\vec{q}}\right]	v_{\vec{q}}(\xN) + \\
		&\phantom{{}=1} \frac{J\hbar^2}{2} \left( 2 \gamma^{\perp}_{ \vec{q}_{\perp}} u_{\vec{q}}^{*}(\xN) + u_{\vec{q}}^{*}(\xN+\A) \right) . \notag
	\end{align}
	\label{eq:scattering.magnons.interfaceconditionstart}%
\end{subequations}
In the remainder of this Appendix, we will obtain the solutions to Eq.~(\ref{eq:scattering.magnons.bulkEquation}) and Eq.~(\ref{eq:scattering.magnons.interfaceconditionstart}).
We first determine bulk solutions that solve Eq.~(\ref{eq:scattering.magnons.bulkEquation}) only, and then we combine the bulk solutions to fulfill the boundary condition of Eq.~(\ref{eq:scattering.magnons.interfaceconditionstart}).

\subsection{Infinite System}
\label{sec:scattering.magnons.infinite}

In bulk antiferromagnets, the system is translationally invariant in all directions. 
In this case, momentum is a good quantum number.
The bulk equations, Eq.~(\ref{eq:scattering.magnons.bulkEquation}), are solved by plane waves,  $u_{\vec{p}} \left( x \right) = u_{\vec{p}} \exp \left( - i p_x x \right)$ and $v_{\vec{p}} \left( x \right) = v_{\vec{p}} \exp \left( i p_x x \right)$.
This solution provides the momentum eigenstates
\begin{subequations}
	\begin{align}
			\beta^{+}_{\vec{p}} &= \sum_{i \in A} \frac{ u_{\vec{p}} }{\sqrt{N/2}} e^{-i \vec{p} \vec{r}_{i}} a_{i} - \sum_{i \in B} \frac{v_{\vec{p}}}{\sqrt{N/2}} e^{i  \vec{p} \vec{r}_{i }} b^{\dag}_{i}, \\
			\beta^{-}_{\vec{p}} &= \sum_{i \in B} \frac{ u_{\vec{p}} }{\sqrt{N/2}} e^{- i \vec{p} \vec{r}_{i}} b_{i} - \sum_{i \in A} \frac{ v_{\vec{p}} }{\sqrt{N/2}} e^{i \vec{p} \vec{r}_{i} } a^{\dag}_{i}  \, .
	\end{align}
	\label{eq:scattering.magnons.def.MomentumCreationAnnihilation}%
\end{subequations}
The Bogoliubov parameters $u_{\vec{p}}$ and $v_{\vec{p}}$ satisfy the normalization condition $u_{\vec{p}}^2-v_{\vec{p}}^2 = 1$. 
From Eq.~(\ref{eq:scattering.magnons.bulkEquation}), we find the dispersion relation for the momentum eigenstates,
\begin{equation}		
	E_{\vec p}	= \hbar^2 \sqrt{\left(3 J + K_z \right)^2 - J^2 \gamma_{\vec{p}}^2 } .
	\label{eq:scattering.magnons.dispersion}
\end{equation}
In an infinite system, the momenta $\vec{p}$ must be real valued since there are no boundaries where evanescent states can originate.

\subsection{Semi-Infinite System}
\label{sec:scattering.magnons.semi}

In a semi-infinite system, there are momentum eigenstates both for complex and real-valued quantum numbers $\vec{p}$. 
When $\vec{p}$ is real, the states represent propagating magnons with momentum $\vec{p}$. 
The momentum eigenstates with complex $\vec{p}$ are evanescent states. 
The broken symmetry at the interface implies that the energy eigenstates are linear combinations of left- and right-going propagating states as well as evanescent states. 
Explicitly, this follows from the boundary condition of Eq.~(\ref{eq:scattering.magnons.interfaceconditionstart}). 
All parts of the energy eigenstates have identical energy in the dispersion of Eq.~(\ref{eq:scattering.magnons.dispersion}) and the same transverse momentum, $\vec{p}_{\perp}$.

We classify the energy eigenstates according to the momentum of the incoming propagating wave, $\vec{p}$. 
The momentum eigenstate with incoming momentum $\vec{p}$ can elastically scatter into three other momentum eigenstates. 
The reflected momentum of one of these states is $\vec{p}^{\R} = (-p_x,\vec{p}_{\perp})$.
The other two states have momenta $\vec{P}$, which satisfy $\gamma_{\vec{P}} = -\gamma_{\vec{p}}$, whereas the transverse momentum remains $\vec{P}_{\perp} = \vec{p}_{\perp}$. 
In total, the energy eigenstates are linear combinations of momentum eigenstates with momenta $\vec{p}$, $\vec{p}^{\R}$, $\vec{P}$ and $\vec{P}^{\R}=(-P_x,\vec{P}_\perp)$. 
For evanescent momentum eigenstates, we only consider states that are localized near the interface.

Three types of energy eigenstates exist. 
There are i) eigenstates containing only plane waves, ii) purely evanescent states, and iii) combinations of plane waves and evanescent waves.
It can be shown that the eigenstates of type i) have energies on the order of the exchange energy.
Such high-energy magnons are not relevant to our discussion of the low-energy physics. 
Therefore, all the states of type i) are disregarded from this point. 
Type ii) and type iii) states have a wide range of energies from the magnon gap and up to the order of the exchange energy. 
We will first write expressions for the magnon eigenstates of type ii) and type iii) that are valid for all energies. 
Later, we will consider the limiting case where the magnon energy is considerably smaller than the exchange energy.

\subsubsection{Surface states}

The evanescent surface states of type ii) are described by\cite{wolfram1969surface}
\begin{subequations}
	\begin{align}
		u_{\vec{q}} \left( x_i \right) &= u_{\vec{q}}^0 e^{ -\Im \left\{q_x\right\} \left( x_i -\xN \right)} + U_{\vec{q}} \left( -1 \right)^{i_x} e^{ -Q_x \left( x_i - \xN \right)},  \\
		v_{\vec{q}} \left( x_i \right) &= v_{\vec{q}}^0 e^{ -\Im \left\{q_x\right\} \left( x_i - \xN \right)} + V_{\vec{q}} \left( -1 \right)^{i_x} e^{ -Q_x \left( x_i - \xN \right) } ,
	\end{align}
	\label{eq:scattering.magnons.surfacestate}%
\end{subequations}
where the longitudinal wave number $q_x$ is purely imaginary. 
The imaginary wave number $\Im \left\{q_x\right\}$ and the real number $Q_x = -i (P_x - \pi / \A)$ are determined by fulfilling both the dispersion relation of Eq.~(\ref{eq:scattering.magnons.dispersion}) and 
\begin{equation}
	E_{\vec{q}}^2 = \left( 3 J + K_z \right)\left( 2 J + K_z \right) \hbar^4 - \gamma^{\perp 2}_{\vec{q}_{\perp}} J^2 \hbar^4 \frac{\left( 3 J + K_z \right)}{\left( 2 J + K_z \right)},
	\label{eq:scattering.magnons.SurfaceEnergy}
\end{equation}
which follows from the boundary conditions of Eq.~(\ref{eq:scattering.magnons.interfaceconditionstart}). 
The coefficients $u_{\vec{q}}^0$, $v_{\vec{q}}^0$, $U_{\vec{q}}$, and $V_{\vec{q}}$ appearing in the surface states of Eq.~(\ref{eq:scattering.magnons.surfacestate}) should be determined from the bulk and boundary conditions of Eq.~(\ref{eq:scattering.magnons.bulkEquation}) and Eq.~(\ref{eq:scattering.magnons.interfaceconditionstart}) and the normalization condition.

For each value of the transverse momenta $q_y$ and $q_z$, there are two surface states; one of type $+$ and one of type $-$.
These states have the same energy and carry opposite angular momentum.

\subsubsection{Delocalized states}

The delocalized states of type iii) are given by the functions
\begin{subequations}
	\begin{align}
		u_{\vec{q}} \! \left( x_i \right) &= u_{\vec{q}}^{\phi} \cos \! \left(q_x \left( x_i {-} \xN \right) \! {-} \phi_{\vec{q}} \right) {+} U_{\vec{q}} \left( -1 \right)^{i_x} \! e^{ -Q_x \left( x_i {-} \xN \right) }, \\
		v_{\vec{q}} \! \left( x_i \right) &= v_{\vec{q}}^{\phi} \cos \! \left(q_x \left( x_i {-} \xN \right) \! {-} \phi_{\vec{q}} \right) {+} V_{\vec{q}} \left( -1 \right)^{i_x} \! e^{ -Q_x \left( x_i {-} \xN \right)} .
	\end{align}
	\label{eq:scattering.magnons.bulkstates}%
\end{subequations}
The real-valued parameter $Q_x$ is governed by $\vec{q}$ via the dispersion relation of Eq.~(\ref{eq:scattering.magnons.dispersion}).
The incoming longitudinal momentum $q_x$ is a free parameter, and the boundary conditions of Eq.~(\ref{eq:scattering.magnons.interfaceconditionstart}) determine the angle $\phi_{\vec{q}}$,
\begin{align}
			&\phantom{{}=1}  \cot\left(q_x \A\right) - \tan \left( \phi_{\vec{q}} \right) = \notag \\
			&\phantom{{}=1} \frac{2 \omega_0 e^{Q_x \A}  - \left( \cosh \left(Q_x \A \right) +\cos \left(q_x \A \right)\right)}{\sin\left(q_x \A \right) \left(2 \omega_0 - e^{Q_x \A} \left( \cosh \left(Q_x \A \right) +\cos \left(q_x \A \right)\right)\right)} ,
	\label{eq:scattering.magnons.phiCondition}
\end{align}
where we have introduced $\omega_0 = 3 + K_z/J$.
The coefficients $u_{\vec{q}}^{\phi}$, $v_{\vec{q}}^{\phi}$, $U_{\vec{q}}$, and $V_{\vec{q}}$ of the delocalized states are governed by the bulk and boundary conditions of Eq.~(\ref{eq:scattering.magnons.bulkEquation}) and Eq.~(\ref{eq:scattering.magnons.interfaceconditionstart}) and the normalization condition.

\subsection{Low-Energy Magnons}
\label{sec:scattering.magnons.lowE}

The magnon wavefunctions given in Sec.~\ref{sec:scattering.magnons.semi} are eigenstates of the
Hamiltonian of Eq.~(\ref{eq:scattering.magnons.Hamiltoniannode}), but the expressions for the magnon currents are complicated. 
In this section, we consider the low-energy limit, where the magnon wavefunctions are simpler. 
We consider a magnon temperature $T_{\Ins}$ that is considerably smaller than the N\'eel temperature, $T_{\Neel}$. 
In this low-energy limit, the magnon energy is much smaller than the exchange energy, $J \hbar^2$. 

\subsubsection{Low-energy momentum eigenstates}

In this section, we study the momentum eigenstates of Sec.~\ref{sec:scattering.magnons.infinite} in the low-energy limit.
When the energy $E_{\vec{p}}$ is considerably smaller than the exchange energy, we see from the dispersion of Eq.~(\ref{eq:scattering.magnons.dispersion}) that $\vec{p}$ satisfies $\gamma_{\vec{p}} \approx \pm 3$.
For low-energy propagating plane waves with momentum $\vec{p}$, $\gamma_{\vec{p}} \approx 3$.  
The low-energy momentum eigenstates with $\gamma_{\vec{P}} \approx - 3$ are staggered evanescent states.

We consider the limit of long wavelength and small anisotropy. 
In other words, we perform expansions in $\abs{\vec{p}} \A$ and $K_z/J$. 
The leading order terms in this expansion for the Bogoliubov parameters, $u_{\vec{p}}$ and $v_{\vec{p}}$, are given by $u_{\vec{p}} = - v_{\vec{p}} = [ 3 \hbar^2 J/( 2 E_{\vec{p}} ) ]^{1/2}$.

The wavefunctions of the staggered evanescent states decay exponentially and oscillate rapidly along the $x$ axis within each sublattice.
Using the dispersion relation of Eq.~(\ref{eq:scattering.magnons.dispersion}), we expand $P_x$ in powers of $\abs{\vec{p}} \A$ and $K_z/J$.
We determine the wavefunctions of the staggered evanescent states in the long wavelength and small anisotropy limit using this expression for $P_x$.
To the leading order in the long wavelength and small anisotropy expansion, the decay length, $1/Q_x$, of the staggered states is $\A / \acosh \left(5\right)$. 

\subsubsection{Low-energy surface states}

We now discuss the surface states of type ii), as given in Eq.~(\ref{eq:scattering.magnons.surfacestate}). 
We expand in the small parameters $\abs{\vec{p}_{\perp}} \A$ and $K_z/J$. The non-staggered part of the wavefunction decays exponentially in the longitudinal ($x$) direction with a decay length $\lambda_{E_{\vec{q}}} = 1/\Im \left\{q_x\right\}$. 
The staggered part has a decay length $1/Q_x$.
Expanding in powers of $\abs{\vec{q}} \A$ and $K_z/J$, we find to leading orders
\begin{subequations}
	\begin{align}
				\Im \left\{q_x\right\} \A &= \frac{1}{\sqrt{6}} \left( \frac{1}{2} \abs{\vec{q}_{\perp}}^2 {\A}^2  + \frac{K_z}{J} \right), \label{eq:scattering.magnons.limKSurface.Imq} \\
				Q_x \A &= \acosh \left(5\right) - \frac{1}{2 \sqrt{6}}\abs{\vec{q}_{\perp}}^2 {\A}^2 \label{eq:scattering.magnons.limKSurface.Q} .
	\end{align}
	\label{eq:scattering.magnons.limKSurface}%
\end{subequations}
Substituting the expression
Eq.~(\ref{eq:scattering.magnons.limKSurface.Imq}) into
Eq.~(\ref{eq:scattering.magnons.dispersion}), we find that to the leading order $E_{\vec{q}} = E_{\vec{q}^{0}}$.
Here, we have defined $\vec{q}^{0} = (0,\vec{q}_{\perp})$.
This means that the low-energy surface state has approximately the same energy as a delocalized state that is uniform along the $x$ direction. 
We will show in this section that we can approximate the low-energy surface states using uniform delocalized states.
The decay length $\lambda_{E_{\vec{q}}} = 6 \sqrt{6} \hbar^4 J^2 \A / E_{\vec{q}}^2$ of the surface state is given by Eq.~(\ref{eq:scattering.magnons.limKSurface.Imq}). 

In the limit of long wavelength and small anisotropy,
\begin{subequations}
	\begin{align}
		U_{\vec{q}} &= V_{\vec{q}} = u_{\vec{q}}^0 \frac{E_{\vec{q}}}{6 \hbar^2 J \left(2 + \sqrt{6} \right)}, \\
		v_{\vec{q}}^0 &= u_{\vec{q}}^0 \left( -1 + \frac{E_{\vec{q} }}{3 \hbar^2 J} \right) .
	\end{align}
	\label{eq:scattering.magnons.limSolutionSurface}%
\end{subequations}
The results in Eq.~(\ref{eq:scattering.magnons.limKSurface}) and Eq.~(\ref{eq:scattering.magnons.limSolutionSurface}) determine the surface state wavefunction completely except for the normalization factor, $u_{\vec{q}}^0$. 
The condition for normalization,
\begin{align}
	\left( u_{\vec{q}}^0 \right)^2 &= \left( u_{\vec{q}^{0}} \right)^2 \frac{2 \Im \left\{q_x\right\} N_x \A }{1 - e^{- 2 \Im \left\{q_x\right\} N_x \A} } ,
	\label{eq:scattering.magnons.SurfaceNormalization}
\end{align}
follows from the commutation relations between the magnon operators, $\alpha^{\pm}_{\vec{q}}$.
The staggered part of the wavefunction does not contribute to Eq.~(\ref{eq:scattering.magnons.SurfaceNormalization}) because the normalization condition is dominated by contributions from the bulk of the AFI and the staggered part of the wavefunction is localized near the interface.
In Eq.~(\ref{eq:scattering.magnons.SurfaceNormalization}), $u_{\vec{q}^{0}}$ is a Bogoliubov parameter of the momentum eigenstate with momentum $\vec{q}^{0}$.
When $N_x \A \ll \lambda_{E_{\vec{q}}}$, the surface state is approximately uniform along the $x$ direction, and the normalization factor $u_{\vec{q}}^0$ is approximately equal to the Bogoliubov parameter $u_{\vec{q}^{0}}$.

In Sec.~\ref{sec:dynamics}, we showed that the relevant surface states to our problem are approximately uniform along the $x$ direction.
These states are approximately equal to uniform delocalized states, meaning that their wavefunctions are approximately equal everywhere in the AFI.
We use the uniform delocalized states to model the behavior of the surface states.
The uniform delocalized states are defined by Eq.~(\ref{eq:scattering.magnons.bulkstates}) for $q_x = 0$ and $\phi_{\vec{q}} = 0$ and satisfy $u^{\phi}_{\vec{q}} = u_{\vec{q}^{0}}$.
Like the surface states, the uniform delocalized states fulfill Eq.~(\ref{eq:scattering.magnons.limSolutionSurface}).

\subsubsection{Low-energy delocalized states}

We now consider the delocalized states of type iii) in the low-energy limit. In general, these states are given by Eq.~(\ref{eq:scattering.magnons.bulkstates}). 
The reflection angle, $\phi_{\vec{q}}$, is determined from Eq.~(\ref{eq:scattering.magnons.phiCondition}). 
In the long-wavelength and low-anisotropy limit, we find that the reflection angle $\phi_{\vec{q}}$ is $ \tan \left(\phi_{\vec{q}}\right) =   1 / \left( q_x \lambda_{E_{\vec{q} }}\right)$.
The smallest possible value of the momentum $q_x$ is on the order of the inverse of the length of the AFI. 
This lower bound provides an upper boundary for the reflection angle, $\phi_{\vec{q}}$, of the order $N_x \A / \lambda_{E_{\vec{q} }}$. 
We have shown in Sec.~\ref{sec:dynamics} that $\lambda_{E_{\vec{q}}}$ is considerably longer than the AFI length, $N_x \A$, for magnon energies, $E_{\vec{q}}$, on the order of the gap, $E_0$. 
Therefore, the reflection angle $\phi_{\vec{q}}$ is small for magnons with energies on the order of the gap.
We have shown in Sec.~\ref{sec:rates} that the reflection angle $\phi_{\vec{q}}$ is also small for thermal magnons with energies that are considerably higher than the gap.
Consequently, we model the behavior of all the states of type iii) using delocalized states with a vanishing reflection angle.

We consider the long-wavelength and small-anisotropy limit. 
For small reflection angles, the delocalized states are described in terms of coefficients that, as the surface states, fulfill Eq.~(\ref{eq:scattering.magnons.limSolutionSurface}). 
However, to the leading order in the long-wavelength and small-anisotropy expansion, the normalization factor differs and is 
\begin{equation}
	u_{\vec{q}}^{\phi} = \sqrt{2}u_{\vec{q}} = \sqrt{ \frac{3 \hbar^2 J}{ E_{\vec{q}}  } }.
	\label{eq:scattering.magnons.BulkNormalization}
\end{equation}
The normalization factor for a state of type iii) is larger than that of a uniform delocalized state by a factor of $\sqrt{2}$. 
The enhancement is caused by constructive interference between the incoming and reflected waves.

\section{Normal Metal Scattering States}
\label{sec:scattering.conduction}

Next, we aim to determine the scattering states in the normal metal.
In the bulk, the Hamiltonian is governed by Eq.~(\ref{eq:dynamics.metalHamiltonian}). 
We will find the energy eigenstates of the conduction electrons when the antiferromagnet is in the classical ground state.
In Sec.~\ref{sec:rates}, we included the coupling of conduction electrons and magnons as a perturbation.
The exchange coupling of Eq.~(\ref{eq:dynamics.InteractionHamiltonian}) results in a spin-dependent potential at the interface that represents the proximity effect.
The total Hamiltonian for the conduction electrons is $H_{\Cond \Tot} = H_{\Cond} + H_{\Int}^{(0)}$, where the interface proximity effect is captured by
\begin{align}
		 H_{\Int}^{(0)} &= -\frac{\hbar^2 J_{\Int} }{4} \sum_{\nearnb{i}{j} \mid i \in A, j \in \Cond} \left( c^{\dag}_{\spinup j} c_{\spinup j} - c^{\dag}_{\spindown j} c_{\spindown j} \right) \notag \\ 
		&+\frac{\hbar^2 J_{\Int} }{4} \sum_{\nearnb{i}{j} \mid i \in B, j \in \Cond} \left( c^{\dag}_{\spinup j} c_{\spinup j} - c^{\dag}_{\spindown j} c_{\spindown j} \right) .
	\label{eq:scattering.conduction.InteractionHamiltonian}
\end{align}

We consider a half-filled system. 
Transport is governed by the states near the Fermi surface. 
In this case, we assume that the scattering states are of the form
\begin{align}
		c^{\dag}_{m \vec{k}} &= \frac{1}{\sqrt{2 M}} \sum_{i} \left( e^{i \vec{k} \cdot \vec{r}_i} + r^{*}_{m \vec{k}} e^{i \vec{k}^{\R} \cdot \vec{r}_i } + r^{* \U}_{m \vec{k}} e^{i \vec{k}^{\U \R} \cdot \vec{r}_i } \right)c^{\dag}_{m i} .
	\label{eq:scattering.conduction.creationannihilationWaveNormal}
\end{align}
The coefficient $r_{m \vec{k}}$ represents specular reflection, and the coefficient $r^{\U}_{m \vec{k}}$ describes Umklapp reflection. 
These coefficients are elements of the $S$ matrix. 
The $S$ matrix is unitary, which ensures the conservation of particle number.
Because the $S$ matrix is unitary, the coefficients $r_{m \vec{k}}$ and $r^{\U}_{m \vec{k}}$ satisfy $r^{*}_{m \vec{k}^{\U}} r^{\U}_{m \vec{k}} + r_{m \vec{k}} r^{\U *}_{m \vec{k}^{\U}}  = 0$ and $1 =  \abs{r_{m \vec{k}}}^2 + \abs{r^{\U}_{m \vec{k}}}^2$. 
These conditions can be used to determine $r_{m \vec{k}}$ and $r^{\U}_{m \vec{k}}$.
The normal metal has the same lengths $M_y \A $ and $M_z \A$ in the $y$ and $z$ directions as the AFI, $M_y=N_y$ and $M_z=N_z$. 
The normal metal extends a finite length $M_x \A$ in the negative $x$ direction from $x = \xN - \D$. 
The number of sites in the normal metal is $M_x M_y M_z$.

The plane wave states, $\sum_{i} \exp^{i \vec{k} \cdot \vec{r}_i} c^{\dag}_{m i}$, are eigenstates of $H_{\Cond}$ except at the interface.
The plane wave states and the scattering states both satisfy the dispersion relation
\begin{equation}
	\varepsilon_{\vec{k}} = -2 t \left[ \cos \left( k_{x} \A \right) + \cos \left( k_{y} \A \right) + \cos \left( k_{z} \A \right)  \right] + 6 t .
	\label{eq:scattering.conduction.dispersion}
\end{equation}
The scattering state $c^{\dag}_{m \vec{k}}$ consists of one incoming plane wave state with respect to the interface and two reflected (normal and Umklapp) plane wave states. 
The wavevector $\vec{k}$ satisfies $\Re \left\{ k_x \right\} \geq 0$ and is therefore incoming.

We require the scattering states of Eq.~(\ref{eq:scattering.conduction.creationannihilationWaveNormal}) to be eigenstates of $H_{\Cond \Tot}$. 
From the Hamiltonian at the sites along the interface, we find the boundary conditions 
\begin{align}
		 &\phantom{{}=1} \frac{\hbar^2 J_{\Int}}{4 t} m \left( e^{-i k^{\U}_x \left( \xN - \D \right)} + r^{*}_{m \vec{k}^{\U}} e^{-i k^{\U \R}_x \left( \xN - \D \right) } \right) \notag \\
		&\phantom{{}=1} = \left(r^{* \U}_{m \vec{k}^{\U}} e^{-i k^{\R}_x \left( \xN - \D + \A\right) } \right), \notag \\
		&\phantom{{}=1} \frac{\hbar^2 J_{\Int}}{4 t} m \left(r^{*\U}_{m \vec{k}} e^{-i k^{\U \R}_x \left( \xN - \D \right) }\right) \notag \\
		&\phantom{{}=1} = \left( e^{-i k_x \left( \xN - \D + \A \right)} + r^{*}_{m \vec{k}} e^{i k^{\R}_x \left( \xN - \D + \A\right) }  \right).
	\label{eq:scattering.conduction.InterfaceCondition}
\end{align}
Defining $\lambda_m = \hbar^2 J_{\Int} m/ \left(4 t\right)$, the solutions to the boundary and unitarity conditions are \cite{cheng2014aspects}
\begin{align}
		r_{m \vec{k}} &= - e^{2 i k_x \left( \xN - \D \right) } \frac{\lambda_m^2 - e^{i \left(k_x - k^{\U}_x \right) \A }}{\lambda_m^2 - e^{-i \left(k_x + k^{\U}_x \right) \A }},\notag \\
		r^{\U}_{m \vec{k}} &= \frac{ 2 i \lambda_m e^{i \left(  k_x +  k^{\U}_x \right)\left( \xN - \D \right) }  \sqrt{ \sin\left( k_x \A \right) \sin\left( k^{\U}_x \A \right) } }{\lambda_m^2 - e^{-i \left(k_x + k^{\U}_x \right) \A }} .
	\label{eq:scattering.conduction.RanChengSolution}
\end{align}
In the limit of weak or strong exchange interaction at the interface, $\abs{\lambda_m} \ll 1$ or $\abs{\lambda_m} \gg 1$, the Umklapp scattering amplitude $r^{\U}_{m \vec{k}}$ and hence the proximity effect are negligible. 
The influence of the antiferromagnet attains its maximum when the exchange interaction is comparable to the next-neighbor hopping parameter. 
The Umklapp scattering amplitude can be related to the spin-mixing conductance\cite{Takei:prb2014}.

\section{Surface Coupling}
\label{sec:scattering.coefficients}

In Appendices \ref{sec:scattering.magnons} and \ref{sec:scattering.conduction}, we have found the energy eigenstates of the noninteracting system of magnons and conduction electrons.
In this section, we determine the dominant interaction term, starting from the interface interaction of Eq.~(\ref{eq:dynamics.InteractionHamiltonian}).
We will arrive at Eqs.~(\ref{eq:dynamics.HamiltonianInterface}) and (\ref{eq:interface.Vcoefficients}).

We perform the Holstein-Primakoff transformation as described in Sec.~\ref{sec:dynamics}, and we expand Eq.~(\ref{eq:dynamics.InteractionHamiltonian}) in powers of magnon operators. 
The zero-order terms have already been accounted for in Appendix \ref{sec:scattering.conduction}, and therefore, we disregard them here.
Disregarding higher-order terms in the magnon operators, we find that
\begin{align}
		H_{\Int} &=  H_{\Int}^{(0)} + \frac{J_{\Int} \hbar^2}{2} \sum_{\nearnb{i}{j} \mid i \in A, j \in \Cond} a_{i} c^{\dag}_{\spindown j} c_{\spinup j} + a^{\dag}_{i} c^{\dag}_{\spinup j} c_{\spindown j} \notag \\
		&+ \frac{J_{\Int} \hbar^2}{2} \sum_{\nearnb{i}{j} \mid i \in B, j \in \Cond} b^{\dag}_{i} c^{\dag}_{\spindown j} c_{\spinup j} + b_{i} c^{\dag}_{\spinup j} c_{\spindown j}.
	\label{eq:scattering.coefficients.InteractionHamiltonian}
\end{align}
We now substitute the delocalized states derived in Appendices \ref{sec:scattering.magnons} and \ref{sec:scattering.conduction} into Eq.~(\ref{eq:scattering.coefficients.InteractionHamiltonian}).
After this substitution, we arrive at Eq.~(\ref{eq:dynamics.HamiltonianInterface}), which defines the interaction amplitudes $V^{\pm}$.

$V^{\pm}$ is determined from the amplitude of the delocalized states at the interface.
The amplitude of a magnon state $\alpha^{+}_{\vec{q}}$ ($\alpha^{-}_{\vec{q}}$) along the interface is given by $u_{\vec{q}} \left( \xN \right)$ on the $A$ ($B$) sublattice and $v_{\vec{q}} \left( \xN \right)$ on sublattice $B$ ($A$).
To the leading order in the magnon exchange energy $\hbar^2 J$, we
find that
\begin{align}
	u_{\vec{q}} \left( \xN \right) &= - v_{\vec{q}} \left( \xN \right) = \frac{\sqrt{2}  u_{\vec{q}} }{ \sqrt{2}^{\delta_{q_x,0}} } = \frac{1}{ \sqrt{2}^{\delta_{q_x,0}} } \sqrt{ \frac{3 \hbar^2 J}{ E_{\vec{q}}  } }.
	\label{eq:scattering.coefficients.magnonAmplitude}
\end{align}
The wavefunction of the conduction electron scattering state $c_{m \vec{k}}$ at the interface has two components.
One component oscillates with momentum $\vec{k}_{\perp}$ and the other oscillates with momentum $\vec{k}^{\U}_{\perp}$ as we move along the interface.
This mixing is caused by the proximity effect. 
The absolute value of the amplitude of each component is constant along the interface.
This absolute value is $C^{\vec{k}}_{m} = \sin\left(k_x \A \right) / (\lambda^2 + 1)$ for the component with momentum $\vec{k}_{\perp}$ and $C^{\vec{k}}_{m} \lambda$ for the component with momentum $\vec{k}^{\U}_{\perp}$. 

Using the amplitudes of the delocalized states, we find to the leading
order that
\begin{widetext}
\begin{align}		
		V^{\pm}_{\vec{q},\vec{K}, \vec{K}'} &= \frac{\hbar^2 J_{\Int} \left( C^{\vec{K}'}_{m} C^{\vec{K}}_{m} \right) }{2 M_x \sqrt{N}}   \sum_{\vec{k},\vec{k}'} \left( e^{\mp i K'_{x} \A} \delta^{\perp}_{\vec{K}', \vec{k}' } - \lambda \delta^{\perp}_{\vec{K}',\vec{k}^{\prime \U} }\right) \left( e^{\pm i K_{x} \A} \delta^{\perp}_{\vec{K},\vec{k}} + \lambda \delta^{\perp}_{\vec{K},\vec{k}^{\U}}\right) \left( u_{\vec{q}} \left( \xN \right) - v_{\vec{q}} \left( \xN \right) \right) \delta^{\perp}_{\pm \vec{q}^{\U},\vec{k}' - \vec{k}}.
	\label{eq:scattering.coefficients.Vcoefficients}
\end{align} 
\end{widetext}
The delta functions $\delta^{\perp}_{\vec{K}', \vec{k}' }$ ensure conservation of momentum in the transverse plane.
Four different combinations of delta functions appear because each of the conduction electron states $c_{m \vec{k}}$ has two components.
These delta functions in turn give rise to two different momentum conditions in Eq.~(\ref{eq:interface.Vcoefficients}).


\begin{thebibliography}{42}%
\makeatletter
\providecommand \@ifxundefined [1]{%
 \@ifx{#1\undefined}
}%
\providecommand \@ifnum [1]{%
 \ifnum #1\expandafter \@firstoftwo
 \else \expandafter \@secondoftwo
 \fi
}%
\providecommand \@ifx [1]{%
 \ifx #1\expandafter \@firstoftwo
 \else \expandafter \@secondoftwo
 \fi
}%
\providecommand \natexlab [1]{#1}%
\providecommand \enquote  [1]{``#1''}%
\providecommand \bibnamefont  [1]{#1}%
\providecommand \bibfnamefont [1]{#1}%
\providecommand \citenamefont [1]{#1}%
\providecommand \href@noop [0]{\@secondoftwo}%
\providecommand \href [0]{\begingroup \@sanitize@url \@href}%
\providecommand \@href[1]{\@@startlink{#1}\@@href}%
\providecommand \@@href[1]{\endgroup#1\@@endlink}%
\providecommand \@sanitize@url [0]{\catcode `\\12\catcode `\$12\catcode
  `\&12\catcode `\#12\catcode `\^12\catcode `\_12\catcode `\%12\relax}%
\providecommand \@@startlink[1]{}%
\providecommand \@@endlink[0]{}%
\providecommand \url  [0]{\begingroup\@sanitize@url \@url }%
\providecommand \@url [1]{\endgroup\@href {#1}{\urlprefix }}%
\providecommand \urlprefix  [0]{URL }%
\providecommand \Eprint [0]{\href }%
\providecommand \doibase [0]{http://dx.doi.org/}%
\providecommand \selectlanguage [0]{\@gobble}%
\providecommand \bibinfo  [0]{\@secondoftwo}%
\providecommand \bibfield  [0]{\@secondoftwo}%
\providecommand \translation [1]{[#1]}%
\providecommand \BibitemOpen [0]{}%
\providecommand \bibitemStop [0]{}%
\providecommand \bibitemNoStop [0]{.\EOS\space}%
\providecommand \EOS [0]{\spacefactor3000\relax}%
\providecommand \BibitemShut  [1]{\csname bibitem#1\endcsname}%
\let\auto@bib@innerbib\@empty
\bibitem [{\citenamefont {Dalfovo}\ \emph {et~al.}(1999)\citenamefont
  {Dalfovo}, \citenamefont {Giorgini}, \citenamefont {Pitaevskii},\ and\
  \citenamefont {Stringari}}]{Dalfovo:rmp1999}%
  \BibitemOpen
  \bibfield  {author} {\bibinfo {author} {\bibfnamefont {F.}~\bibnamefont
  {Dalfovo}}, \bibinfo {author} {\bibfnamefont {S.}~\bibnamefont {Giorgini}},
  \bibinfo {author} {\bibfnamefont {L.~P.}\ \bibnamefont {Pitaevskii}}, \ and\
  \bibinfo {author} {\bibfnamefont {S.}~\bibnamefont {Stringari}},\ }\href
  {http://link.aps.org/abstract/RMP/v71/p463} {\bibfield  {journal} {\bibinfo
  {journal} {Rev. Mod. Phys.}\ }\textbf {\bibinfo {volume} {71}},\ \bibinfo
  {pages} {463} (\bibinfo {year} {1999})}\BibitemShut {NoStop}%
\bibitem [{\citenamefont {Nikuni}\ \emph {et~al.}(2000)\citenamefont {Nikuni},
  \citenamefont {Oshikawa}, \citenamefont {Oosawa},\ and\ \citenamefont
  {Tanaka}}]{Nikuni:prl2000}%
  \BibitemOpen
  \bibfield  {author} {\bibinfo {author} {\bibfnamefont {T.}~\bibnamefont
  {Nikuni}}, \bibinfo {author} {\bibfnamefont {M.}~\bibnamefont {Oshikawa}},
  \bibinfo {author} {\bibfnamefont {A.}~\bibnamefont {Oosawa}}, \ and\ \bibinfo
  {author} {\bibfnamefont {H.}~\bibnamefont {Tanaka}},\ }\href
  {http://link.aps.org/doi/10.1103/PhysRevLett.84.5868} {\bibfield  {journal}
  {\bibinfo  {journal} {Phys. Rev. Lett.}\ }\textbf {\bibinfo {volume} {84}},\
  \bibinfo {pages} {5868} (\bibinfo {year} {2000})}\BibitemShut {NoStop}%
\bibitem [{\citenamefont {Ruegg}\ \emph {et~al.}(2003)\citenamefont {Ruegg},
  \citenamefont {Cavadini}, \citenamefont {Furrer}, \citenamefont {Gudel},
  \citenamefont {Kramer}, \citenamefont {Mutka}, \citenamefont {Wildes},
  \citenamefont {Habicht},\ and\ \citenamefont
  {Vorderwisch}}]{Ruegg:nature2003}%
  \BibitemOpen
  \bibfield  {author} {\bibinfo {author} {\bibfnamefont {C.}~\bibnamefont
  {Ruegg}}, \bibinfo {author} {\bibfnamefont {N.}~\bibnamefont {Cavadini}},
  \bibinfo {author} {\bibfnamefont {A.}~\bibnamefont {Furrer}}, \bibinfo
  {author} {\bibfnamefont {H.~U.}\ \bibnamefont {Gudel}}, \bibinfo {author}
  {\bibfnamefont {K.}~\bibnamefont {Kramer}}, \bibinfo {author} {\bibfnamefont
  {H.}~\bibnamefont {Mutka}}, \bibinfo {author} {\bibfnamefont
  {A.}~\bibnamefont {Wildes}}, \bibinfo {author} {\bibfnamefont
  {K.}~\bibnamefont {Habicht}}, \ and\ \bibinfo {author} {\bibfnamefont
  {P.}~\bibnamefont {Vorderwisch}},\ }\href
  {http://dx.doi.org/10.1038/nature01617} {\bibfield  {journal} {\bibinfo
  {journal} {Nature (London)}\ }\textbf {\bibinfo {volume} {423}},\ \bibinfo
  {pages} {62} (\bibinfo {year} {2003})}\BibitemShut {NoStop}%
\bibitem [{\citenamefont {Radu}\ \emph {et~al.}(2005)\citenamefont {Radu},
  \citenamefont {Wilhelm}, \citenamefont {Yushankhai}, \citenamefont
  {Kovrizhin}, \citenamefont {Coldea}, \citenamefont {Tylczynski},
  \citenamefont {Lühmann},\ and\ \citenamefont {Steglich}}]{Radu:prl2005}%
  \BibitemOpen
  \bibfield  {author} {\bibinfo {author} {\bibfnamefont {T.}~\bibnamefont
  {Radu}}, \bibinfo {author} {\bibfnamefont {H.}~\bibnamefont {Wilhelm}},
  \bibinfo {author} {\bibfnamefont {V.}~\bibnamefont {Yushankhai}}, \bibinfo
  {author} {\bibfnamefont {D.}~\bibnamefont {Kovrizhin}}, \bibinfo {author}
  {\bibfnamefont {R.}~\bibnamefont {Coldea}}, \bibinfo {author} {\bibfnamefont
  {Z.}~\bibnamefont {Tylczynski}}, \bibinfo {author} {\bibfnamefont
  {T.}~\bibnamefont {Lühmann}}, \ and\ \bibinfo {author} {\bibfnamefont
  {F.}~\bibnamefont {Steglich}},\ }\href
  {http://link.aps.org/doi/10.1103/PhysRevLett.95.127202} {\bibfield  {journal}
  {\bibinfo  {journal} {Phys. Rev. Lett.}\ }\textbf {\bibinfo {volume} {95}},\
  \bibinfo {pages} {127202} (\bibinfo {year} {2005})}\BibitemShut {NoStop}%
\bibitem [{\citenamefont {Demokritov}\ \emph {et~al.}(2006)\citenamefont
  {Demokritov}, \citenamefont {Demidov}, \citenamefont {Dzyapko}, \citenamefont
  {Melkov}, \citenamefont {Serga}, \citenamefont {Hillebrands},\ and\
  \citenamefont {Slavin}}]{Demokritov:Nature2006}%
  \BibitemOpen
  \bibfield  {author} {\bibinfo {author} {\bibfnamefont {S.~O.}\ \bibnamefont
  {Demokritov}}, \bibinfo {author} {\bibfnamefont {V.~E.}\ \bibnamefont
  {Demidov}}, \bibinfo {author} {\bibfnamefont {O.}~\bibnamefont {Dzyapko}},
  \bibinfo {author} {\bibfnamefont {G.~A.}\ \bibnamefont {Melkov}}, \bibinfo
  {author} {\bibfnamefont {A.~A.}\ \bibnamefont {Serga}}, \bibinfo {author}
  {\bibfnamefont {B.}~\bibnamefont {Hillebrands}}, \ and\ \bibinfo {author}
  {\bibfnamefont {A.~N.}\ \bibnamefont {Slavin}},\ }\href
  {http://dx.doi.org/10.1038/nature05117} {\bibfield  {journal} {\bibinfo
  {journal} {Nature (London)}\ }\textbf {\bibinfo {volume} {443}},\ \bibinfo
  {pages} {430} (\bibinfo {year} {2006})}\BibitemShut {NoStop}%
\bibitem [{\citenamefont {Kasprzak}\ \emph {et~al.}(2006)\citenamefont
  {Kasprzak}, \citenamefont {Richard}, \citenamefont {Kundermann},
  \citenamefont {Baas}, \citenamefont {Jeambrun}, \citenamefont {Keeling},
  \citenamefont {Marchetti}, \citenamefont {Szymanska}, \citenamefont {Andre},
  \citenamefont {Staehli}, \citenamefont {Savona}, \citenamefont {Littlewood},
  \citenamefont {Deveaud},\ and\ \citenamefont {Dang}}]{Kasprzak:nature2006}%
  \BibitemOpen
  \bibfield  {author} {\bibinfo {author} {\bibfnamefont {J.}~\bibnamefont
  {Kasprzak}}, \bibinfo {author} {\bibfnamefont {M.}~\bibnamefont {Richard}},
  \bibinfo {author} {\bibfnamefont {S.}~\bibnamefont {Kundermann}}, \bibinfo
  {author} {\bibfnamefont {A.}~\bibnamefont {Baas}}, \bibinfo {author}
  {\bibfnamefont {P.}~\bibnamefont {Jeambrun}}, \bibinfo {author}
  {\bibfnamefont {J.~M.~J.}\ \bibnamefont {Keeling}}, \bibinfo {author}
  {\bibfnamefont {F.~M.}\ \bibnamefont {Marchetti}}, \bibinfo {author}
  {\bibfnamefont {M.~H.}\ \bibnamefont {Szymanska}}, \bibinfo {author}
  {\bibfnamefont {R.}~\bibnamefont {Andre}}, \bibinfo {author} {\bibfnamefont
  {J.~L.}\ \bibnamefont {Staehli}}, \bibinfo {author} {\bibfnamefont
  {V.}~\bibnamefont {Savona}}, \bibinfo {author} {\bibfnamefont {P.~B.}\
  \bibnamefont {Littlewood}}, \bibinfo {author} {\bibfnamefont
  {B.}~\bibnamefont {Deveaud}}, \ and\ \bibinfo {author} {\bibfnamefont
  {L.~S.}\ \bibnamefont {Dang}},\ }\href {\doibase
  http://www.nature.com/nature/journal/v443/n7110/suppinfo/nature05131_S1.html}
  {\bibfield  {journal} {\bibinfo  {journal} {Nature (London)}\ }\textbf
  {\bibinfo {volume} {443}},\ \bibinfo {pages} {409} (\bibinfo {year}
  {2006})}\BibitemShut {NoStop}%
\bibitem [{\citenamefont {Balili}\ \emph {et~al.}(2007)\citenamefont {Balili},
  \citenamefont {Hartwell}, \citenamefont {Snoke}, \citenamefont {Pfeiffer},\
  and\ \citenamefont {West}}]{Balili:science2007}%
  \BibitemOpen
  \bibfield  {author} {\bibinfo {author} {\bibfnamefont {R.}~\bibnamefont
  {Balili}}, \bibinfo {author} {\bibfnamefont {V.}~\bibnamefont {Hartwell}},
  \bibinfo {author} {\bibfnamefont {D.}~\bibnamefont {Snoke}}, \bibinfo
  {author} {\bibfnamefont {L.}~\bibnamefont {Pfeiffer}}, \ and\ \bibinfo
  {author} {\bibfnamefont {K.}~\bibnamefont {West}},\ }\href {\doibase
  10.1126/science.1140990} {\bibfield  {journal} {\bibinfo  {journal}
  {Science}\ }\textbf {\bibinfo {volume} {316}},\ \bibinfo {pages} {1007}
  (\bibinfo {year} {2007})}\BibitemShut {NoStop}%
\bibitem [{\citenamefont {Klaers}\ \emph {et~al.}(2010)\citenamefont {Klaers},
  \citenamefont {Schmitt}, \citenamefont {Vewinger},\ and\ \citenamefont
  {Weitz}}]{Klaers:nature2010}%
  \BibitemOpen
  \bibfield  {author} {\bibinfo {author} {\bibfnamefont {J.}~\bibnamefont
  {Klaers}}, \bibinfo {author} {\bibfnamefont {J.}~\bibnamefont {Schmitt}},
  \bibinfo {author} {\bibfnamefont {F.}~\bibnamefont {Vewinger}}, \ and\
  \bibinfo {author} {\bibfnamefont {M.}~\bibnamefont {Weitz}},\ }\href
  {http://dx.doi.org/10.1038/nature09567} {\bibfield  {journal} {\bibinfo
  {journal} {Nature (London)}\ }\textbf {\bibinfo {volume} {468}},\ \bibinfo
  {pages} {545} (\bibinfo {year} {2010})}\BibitemShut {NoStop}%
\bibitem [{\citenamefont {Bunkov}\ \emph {et~al.}(2011)\citenamefont {Bunkov},
  \citenamefont {Alakshin}, \citenamefont {Gazizulin}, \citenamefont
  {Klochkov}, \citenamefont {Kuzmin}, \citenamefont {Safin},\ and\
  \citenamefont {Tagirov}}]{Bunkov:JETP2011}%
  \BibitemOpen
  \bibfield  {author} {\bibinfo {author} {\bibfnamefont {Y.~M.}\ \bibnamefont
  {Bunkov}}, \bibinfo {author} {\bibfnamefont {E.~M.}\ \bibnamefont
  {Alakshin}}, \bibinfo {author} {\bibfnamefont {R.~R.}\ \bibnamefont
  {Gazizulin}}, \bibinfo {author} {\bibfnamefont {A.~V.}\ \bibnamefont
  {Klochkov}}, \bibinfo {author} {\bibfnamefont {V.~V.}\ \bibnamefont
  {Kuzmin}}, \bibinfo {author} {\bibfnamefont {T.~R.}\ \bibnamefont {Safin}}, \
  and\ \bibinfo {author} {\bibfnamefont {M.~S.}\ \bibnamefont {Tagirov}},\
  }\href {\doibase 10.1134/S0021364011130066} {\bibfield  {journal} {\bibinfo
  {journal} {JETP Lett.}\ }\textbf {\bibinfo {volume} {94}},\ \bibinfo {pages}
  {68} (\bibinfo {year} {2011})}\BibitemShut {NoStop}%
\bibitem [{\citenamefont {Sonin}(1978)}]{Sonin:ssc1978}%
  \BibitemOpen
  \bibfield  {author} {\bibinfo {author} {\bibfnamefont {E.~B.}\ \bibnamefont
  {Sonin}},\ }\href
  {http://www.sciencedirect.com/science/article/pii/0038109878902259}
  {\bibfield  {journal} {\bibinfo  {journal} {Solid State Commun.}\ }\textbf
  {\bibinfo {volume} {25}},\ \bibinfo {pages} {253} (\bibinfo {year}
  {1978})}\BibitemShut {NoStop}%
\bibitem [{\citenamefont {Sonin}(2010)}]{Sonin:adp2010}%
  \BibitemOpen
  \bibfield  {author} {\bibinfo {author} {\bibfnamefont {E.~B.}\ \bibnamefont
  {Sonin}},\ }\href {http://dx.doi.org/10.1080/00018731003739943} {\bibfield
  {journal} {\bibinfo  {journal} {Adv. Phys.}\ }\textbf {\bibinfo {volume}
  {59}},\ \bibinfo {pages} {181} (\bibinfo {year} {2010})}\BibitemShut
  {NoStop}%
\bibitem [{\citenamefont {Bunkov}\ \emph {et~al.}(2012)\citenamefont {Bunkov},
  \citenamefont {Alakshin}, \citenamefont {Gazizulin}, \citenamefont
  {Klochkov}, \citenamefont {Kuzmin}, \citenamefont {L’vov},\ and\
  \citenamefont {Tagirov}}]{Bunkov:prl2012}%
  \BibitemOpen
  \bibfield  {author} {\bibinfo {author} {\bibfnamefont {Y.~M.}\ \bibnamefont
  {Bunkov}}, \bibinfo {author} {\bibfnamefont {E.~M.}\ \bibnamefont
  {Alakshin}}, \bibinfo {author} {\bibfnamefont {R.~R.}\ \bibnamefont
  {Gazizulin}}, \bibinfo {author} {\bibfnamefont {A.~V.}\ \bibnamefont
  {Klochkov}}, \bibinfo {author} {\bibfnamefont {V.~V.}\ \bibnamefont
  {Kuzmin}}, \bibinfo {author} {\bibfnamefont {V.~S.}\ \bibnamefont {L’vov}},
  \ and\ \bibinfo {author} {\bibfnamefont {M.~S.}\ \bibnamefont {Tagirov}},\
  }\href {http://link.aps.org/doi/10.1103/PhysRevLett.108.177002} {\bibfield
  {journal} {\bibinfo  {journal} {Phys. Rev. Lett.}\ }\textbf {\bibinfo
  {volume} {108}},\ \bibinfo {pages} {177002} (\bibinfo {year}
  {2012})}\BibitemShut {NoStop}%
\bibitem [{\citenamefont {Chen}\ and\ \citenamefont
  {Sigrist}(2014)}]{Chen:prb2014}%
  \BibitemOpen
  \bibfield  {author} {\bibinfo {author} {\bibfnamefont {W.}~\bibnamefont
  {Chen}}\ and\ \bibinfo {author} {\bibfnamefont {M.}~\bibnamefont {Sigrist}},\
  }\href {http://link.aps.org/doi/10.1103/PhysRevB.89.024511} {\bibfield
  {journal} {\bibinfo  {journal} {Phys. Rev. B}\ }\textbf {\bibinfo {volume}
  {89}},\ \bibinfo {pages} {024511} (\bibinfo {year} {2014})}\BibitemShut
  {NoStop}%
\bibitem [{\citenamefont {Takei}\ and\ \citenamefont
  {Tserkovnyak}(2014)}]{Takei:prl2014}%
  \BibitemOpen
  \bibfield  {author} {\bibinfo {author} {\bibfnamefont {S.}~\bibnamefont
  {Takei}}\ and\ \bibinfo {author} {\bibfnamefont {Y.}~\bibnamefont
  {Tserkovnyak}},\ }\href
  {http://link.aps.org/doi/10.1103/PhysRevLett.112.227201} {\bibfield
  {journal} {\bibinfo  {journal} {Phys. Rev. Lett.}\ }\textbf {\bibinfo
  {volume} {112}},\ \bibinfo {pages} {227201} (\bibinfo {year}
  {2014})}\BibitemShut {NoStop}%
\bibitem [{\citenamefont {Takei}\ \emph {et~al.}(2014)\citenamefont {Takei},
  \citenamefont {Halperin}, \citenamefont {Yacoby},\ and\ \citenamefont
  {Tserkovnyak}}]{Takei:prb2014}%
  \BibitemOpen
  \bibfield  {author} {\bibinfo {author} {\bibfnamefont {S.}~\bibnamefont
  {Takei}}, \bibinfo {author} {\bibfnamefont {B.~I.}\ \bibnamefont {Halperin}},
  \bibinfo {author} {\bibfnamefont {A.}~\bibnamefont {Yacoby}}, \ and\ \bibinfo
  {author} {\bibfnamefont {Y.}~\bibnamefont {Tserkovnyak}},\ }\href
  {http://link.aps.org/doi/10.1103/PhysRevB.90.094408} {\bibfield  {journal}
  {\bibinfo  {journal} {Phys. Rev. B}\ }\textbf {\bibinfo {volume} {90}},\
  \bibinfo {pages} {094408} (\bibinfo {year} {2014})}\BibitemShut {NoStop}%
\bibitem [{\citenamefont {Skarsvåg}\ \emph {et~al.}(2015)\citenamefont
  {Skarsvåg}, \citenamefont {Holmqvist},\ and\ \citenamefont
  {Brataas}}]{Skarsvag:prl2015}%
  \BibitemOpen
  \bibfield  {author} {\bibinfo {author} {\bibfnamefont {H.}~\bibnamefont
  {Skarsvåg}}, \bibinfo {author} {\bibfnamefont {C.}~\bibnamefont
  {Holmqvist}}, \ and\ \bibinfo {author} {\bibfnamefont {A.}~\bibnamefont
  {Brataas}},\ }\href {http://link.aps.org/doi/10.1103/PhysRevLett.115.237201}
  {\bibfield  {journal} {\bibinfo  {journal} {Phys. Rev. Lett.}\ }\textbf
  {\bibinfo {volume} {115}},\ \bibinfo {pages} {237201} (\bibinfo {year}
  {2015})}\BibitemShut {NoStop}%
\bibitem [{\citenamefont {Flebus}\ \emph {et~al.}(2016)\citenamefont {Flebus},
  \citenamefont {Bender}, \citenamefont {Tserkovnyak},\ and\ \citenamefont
  {Duine}}]{Flebus:prl2016}%
  \BibitemOpen
  \bibfield  {author} {\bibinfo {author} {\bibfnamefont {B.}~\bibnamefont
  {Flebus}}, \bibinfo {author} {\bibfnamefont {S.~A.}\ \bibnamefont {Bender}},
  \bibinfo {author} {\bibfnamefont {Y.}~\bibnamefont {Tserkovnyak}}, \ and\
  \bibinfo {author} {\bibfnamefont {R.~A.}\ \bibnamefont {Duine}},\ }\href
  {http://link.aps.org/doi/10.1103/PhysRevLett.116.117201} {\bibfield
  {journal} {\bibinfo  {journal} {Phys. Rev. Lett.}\ }\textbf {\bibinfo
  {volume} {116}},\ \bibinfo {pages} {117201} (\bibinfo {year}
  {2016})}\BibitemShut {NoStop}%
\bibitem [{\citenamefont {Takei}\ and\ \citenamefont
  {Tserkovnyak}(2015)}]{Takei:prl2016}%
  \BibitemOpen
  \bibfield  {author} {\bibinfo {author} {\bibfnamefont {S.}~\bibnamefont
  {Takei}}\ and\ \bibinfo {author} {\bibfnamefont {Y.}~\bibnamefont
  {Tserkovnyak}},\ }\href
  {http://link.aps.org/doi/10.1103/PhysRevLett.115.156604} {\bibfield
  {journal} {\bibinfo  {journal} {Phys. Rev. Lett.}\ }\textbf {\bibinfo
  {volume} {115}},\ \bibinfo {pages} {156604} (\bibinfo {year}
  {2015})}\BibitemShut {NoStop}%
\bibitem [{\citenamefont {Sun}\ \emph {et~al.}(2016)\citenamefont {Sun},
  \citenamefont {Nattermann},\ and\ \citenamefont {Pokrovsky}}]{Sun:prl2016}%
  \BibitemOpen
  \bibfield  {author} {\bibinfo {author} {\bibfnamefont {C.}~\bibnamefont
  {Sun}}, \bibinfo {author} {\bibfnamefont {T.}~\bibnamefont {Nattermann}}, \
  and\ \bibinfo {author} {\bibfnamefont {V.~L.}\ \bibnamefont {Pokrovsky}},\
  }\href {http://link.aps.org/doi/10.1103/PhysRevLett.116.257205} {\bibfield
  {journal} {\bibinfo  {journal} {Phys. Rev. Lett.}\ }\textbf {\bibinfo
  {volume} {116}},\ \bibinfo {pages} {257205} (\bibinfo {year}
  {2016})}\BibitemShut {NoStop}%
\bibitem [{\citenamefont {Cheng}\ \emph {et~al.}(2014)\citenamefont {Cheng},
  \citenamefont {Xiao}, \citenamefont {Niu},\ and\ \citenamefont
  {Brataas}}]{Cheng:prl2014}%
  \BibitemOpen
  \bibfield  {author} {\bibinfo {author} {\bibfnamefont {R.}~\bibnamefont
  {Cheng}}, \bibinfo {author} {\bibfnamefont {J.}~\bibnamefont {Xiao}},
  \bibinfo {author} {\bibfnamefont {Q.}~\bibnamefont {Niu}}, \ and\ \bibinfo
  {author} {\bibfnamefont {A.}~\bibnamefont {Brataas}},\ }\href@noop {}
  {\bibfield  {journal} {\bibinfo  {journal} {Phys. Rev. Lett.}\ }\textbf
  {\bibinfo {volume} {113}},\ \bibinfo {pages} {057601} (\bibinfo {year}
  {2014})}\BibitemShut {NoStop}%
\bibitem [{\citenamefont {Cheng}\ \emph {et~al.}(2016)\citenamefont {Cheng},
  \citenamefont {Xiao},\ and\ \citenamefont {Brataas}}]{Cheng:prl2016}%
  \BibitemOpen
  \bibfield  {author} {\bibinfo {author} {\bibfnamefont {R.}~\bibnamefont
  {Cheng}}, \bibinfo {author} {\bibfnamefont {D.}~\bibnamefont {Xiao}}, \ and\
  \bibinfo {author} {\bibfnamefont {A.}~\bibnamefont {Brataas}},\ }\href@noop
  {} {\bibfield  {journal} {\bibinfo  {journal} {Phys. Rev. Lett.}\ }\textbf
  {\bibinfo {volume} {116}},\ \bibinfo {pages} {207603} (\bibinfo {year}
  {2016})}\BibitemShut {NoStop}%
\bibitem [{\citenamefont {Tveten}\ \emph {et~al.}(2013)\citenamefont {Tveten},
  \citenamefont {Qaiumzadeh}, \citenamefont {Tretiakov},\ and\ \citenamefont
  {Brataas}}]{Tveten:prl2013}%
  \BibitemOpen
  \bibfield  {author} {\bibinfo {author} {\bibfnamefont {E.}~\bibnamefont
  {Tveten}}, \bibinfo {author} {\bibfnamefont {A.}~\bibnamefont {Qaiumzadeh}},
  \bibinfo {author} {\bibfnamefont {O.~A.}\ \bibnamefont {Tretiakov}}, \ and\
  \bibinfo {author} {\bibfnamefont {A.}~\bibnamefont {Brataas}},\ }\href@noop
  {} {\bibfield  {journal} {\bibinfo  {journal} {Phys. Rev. Lett.}\ }\textbf
  {\bibinfo {volume} {110}},\ \bibinfo {pages} {127208} (\bibinfo {year}
  {2013})}\BibitemShut {NoStop}%
\bibitem [{\citenamefont {Tveten}\ \emph {et~al.}(2014)\citenamefont {Tveten},
  \citenamefont {Qaiumzadeh},\ and\ \citenamefont {Brataas}}]{Tveten:prl2014}%
  \BibitemOpen
  \bibfield  {author} {\bibinfo {author} {\bibfnamefont {E.}~\bibnamefont
  {Tveten}}, \bibinfo {author} {\bibfnamefont {A.}~\bibnamefont {Qaiumzadeh}},
  \ and\ \bibinfo {author} {\bibfnamefont {A.}~\bibnamefont {Brataas}},\
  }\href@noop {} {\bibfield  {journal} {\bibinfo  {journal} {Phys. Rev. Lett.}\
  }\textbf {\bibinfo {volume} {112}},\ \bibinfo {pages} {147204} (\bibinfo
  {year} {2014})}\BibitemShut {NoStop}%
\bibitem [{\citenamefont {Wang}\ \emph {et~al.}(2014)\citenamefont {Wang},
  \citenamefont {Du}, \citenamefont {Hammel},\ and\ \citenamefont
  {Yang}}]{Wang:prl2014}%
  \BibitemOpen
  \bibfield  {author} {\bibinfo {author} {\bibfnamefont {H.}~\bibnamefont
  {Wang}}, \bibinfo {author} {\bibfnamefont {C.}~\bibnamefont {Du}}, \bibinfo
  {author} {\bibfnamefont {P.~C.}\ \bibnamefont {Hammel}}, \ and\ \bibinfo
  {author} {\bibfnamefont {F.}~\bibnamefont {Yang}},\ }\href@noop {} {\bibfield
   {journal} {\bibinfo  {journal} {Phys. Rev. Lett.}\ }\textbf {\bibinfo
  {volume} {113}},\ \bibinfo {pages} {097202} (\bibinfo {year}
  {2014})}\BibitemShut {NoStop}%
\bibitem [{\citenamefont {Wang}\ \emph {et~al.}(2015)\citenamefont {Wang},
  \citenamefont {Du}, \citenamefont {Hammel},\ and\ \citenamefont
  {Yang}}]{Wang:prb2015}%
  \BibitemOpen
  \bibfield  {author} {\bibinfo {author} {\bibfnamefont {H.}~\bibnamefont
  {Wang}}, \bibinfo {author} {\bibfnamefont {C.}~\bibnamefont {Du}}, \bibinfo
  {author} {\bibfnamefont {P.~C.}\ \bibnamefont {Hammel}}, \ and\ \bibinfo
  {author} {\bibfnamefont {F.}~\bibnamefont {Yang}},\ }\href@noop {} {\bibfield
   {journal} {\bibinfo  {journal} {Phys. Rev. B}\ }\textbf {\bibinfo {volume}
  {91}},\ \bibinfo {pages} {220410(R)} (\bibinfo {year} {2015})}\BibitemShut
  {NoStop}%
\bibitem [{\citenamefont {Takei}\ \emph {et~al.}(2015)\citenamefont {Takei},
  \citenamefont {Moriyama}, \citenamefont {Onon},\ and\ \citenamefont
  {Tserkovnyak}}]{Takei:prb2015}%
  \BibitemOpen
  \bibfield  {author} {\bibinfo {author} {\bibfnamefont {S.}~\bibnamefont
  {Takei}}, \bibinfo {author} {\bibfnamefont {T.}~\bibnamefont {Moriyama}},
  \bibinfo {author} {\bibfnamefont {T.}~\bibnamefont {Onon}}, \ and\ \bibinfo
  {author} {\bibfnamefont {Y.}~\bibnamefont {Tserkovnyak}},\ }\href@noop {}
  {\bibfield  {journal} {\bibinfo  {journal} {Phys. Rev. B}\ }\textbf {\bibinfo
  {volume} {92}},\ \bibinfo {pages} {020409(R)} (\bibinfo {year}
  {2015})}\BibitemShut {NoStop}%
\bibitem [{\citenamefont {Moriyama}\ \emph {et~al.}(2015)\citenamefont
  {Moriyama}, \citenamefont {Takei}, \citenamefont {Nagata}, \citenamefont
  {Yoshimura}, \citenamefont {Matsuzaki},\ and\ \citenamefont
  {Terashima}}]{Moriyama:apl2015}%
  \BibitemOpen
  \bibfield  {author} {\bibinfo {author} {\bibfnamefont {T.}~\bibnamefont
  {Moriyama}}, \bibinfo {author} {\bibfnamefont {S.}~\bibnamefont {Takei}},
  \bibinfo {author} {\bibfnamefont {M.}~\bibnamefont {Nagata}}, \bibinfo
  {author} {\bibfnamefont {Y.}~\bibnamefont {Yoshimura}}, \bibinfo {author}
  {\bibfnamefont {N.}~\bibnamefont {Matsuzaki}}, \ and\ \bibinfo {author}
  {\bibfnamefont {T.}~\bibnamefont {Terashima}},\ }\href@noop {} {\bibfield
  {journal} {\bibinfo  {journal} {Appl. Phys. Lett.}\ }\textbf {\bibinfo
  {volume} {106}},\ \bibinfo {pages} {162406} (\bibinfo {year}
  {2015})}\BibitemShut {NoStop}%
\bibitem [{\citenamefont {Tveten}\ \emph {et~al.}(2016)\citenamefont {Tveten},
  \citenamefont {M\"{u}ller}, \citenamefont {Linder},\ and\ \citenamefont
  {Brataas}}]{Tveten:prb2016}%
  \BibitemOpen
  \bibfield  {author} {\bibinfo {author} {\bibfnamefont {E.~G.}\ \bibnamefont
  {Tveten}}, \bibinfo {author} {\bibfnamefont {T.}~\bibnamefont {M\"{u}ller}},
  \bibinfo {author} {\bibfnamefont {J.}~\bibnamefont {Linder}}, \ and\ \bibinfo
  {author} {\bibfnamefont {A.}~\bibnamefont {Brataas}},\ }\href@noop {}
  {\bibfield  {journal} {\bibinfo  {journal} {Phys. Rev. B}\ }\textbf {\bibinfo
  {volume} {93}},\ \bibinfo {pages} {104408} (\bibinfo {year}
  {2016})}\BibitemShut {NoStop}%
\bibitem [{\citenamefont {Bender}\ \emph {et~al.}(2012)\citenamefont {Bender},
  \citenamefont {Duine},\ and\ \citenamefont
  {Tserkovnyak}}]{PhysRevLett.108.246601}%
  \BibitemOpen
  \bibfield  {author} {\bibinfo {author} {\bibfnamefont {S.~A.}\ \bibnamefont
  {Bender}}, \bibinfo {author} {\bibfnamefont {R.~A.}\ \bibnamefont {Duine}}, \
  and\ \bibinfo {author} {\bibfnamefont {Y.}~\bibnamefont {Tserkovnyak}},\
  }\href {\doibase 10.1103/PhysRevLett.108.246601} {\bibfield  {journal}
  {\bibinfo  {journal} {Phys. Rev. Lett.}\ }\textbf {\bibinfo {volume} {108}},\
  \bibinfo {pages} {246601} (\bibinfo {year} {2012})}\BibitemShut {NoStop}%
\bibitem [{\citenamefont {Bender}\ \emph {et~al.}(2014)\citenamefont {Bender},
  \citenamefont {Duine}, \citenamefont {Brataas},\ and\ \citenamefont
  {Tserkovnyak}}]{bender2014dynamic}%
  \BibitemOpen
  \bibfield  {author} {\bibinfo {author} {\bibfnamefont {S.~A.}\ \bibnamefont
  {Bender}}, \bibinfo {author} {\bibfnamefont {R.~A.}\ \bibnamefont {Duine}},
  \bibinfo {author} {\bibfnamefont {A.}~\bibnamefont {Brataas}}, \ and\
  \bibinfo {author} {\bibfnamefont {Y.}~\bibnamefont {Tserkovnyak}},\
  }\href@noop {} {\bibfield  {journal} {\bibinfo  {journal} {Phys. Rev. B}\
  }\textbf {\bibinfo {volume} {90}},\ \bibinfo {pages} {094409} (\bibinfo
  {year} {2014})}\BibitemShut {NoStop}%
\bibitem [{\citenamefont {Gomonay}\ and\ \citenamefont
  {Loktev}(2010)}]{Gomonay:prb2010}%
  \BibitemOpen
  \bibfield  {author} {\bibinfo {author} {\bibfnamefont {H.~V.}\ \bibnamefont
  {Gomonay}}\ and\ \bibinfo {author} {\bibfnamefont {V.~M.}\ \bibnamefont
  {Loktev}},\ }\href@noop {} {\bibfield  {journal} {\bibinfo  {journal} {Phys.
  Rev. B}\ }\textbf {\bibinfo {volume} {81}},\ \bibinfo {pages} {144427}
  (\bibinfo {year} {2010})}\BibitemShut {NoStop}%
\bibitem [{\citenamefont {Sinova}\ \emph {et~al.}(2015)\citenamefont {Sinova},
  \citenamefont {Valenzuela}, \citenamefont {Wunderlich}, \citenamefont
  {Back},\ and\ \citenamefont {Jungwirth}}]{RevModPhys.87.1213}%
  \BibitemOpen
  \bibfield  {author} {\bibinfo {author} {\bibfnamefont {J.}~\bibnamefont
  {Sinova}}, \bibinfo {author} {\bibfnamefont {S.~O.}\ \bibnamefont
  {Valenzuela}}, \bibinfo {author} {\bibfnamefont {J.}~\bibnamefont
  {Wunderlich}}, \bibinfo {author} {\bibfnamefont {C.~H.}\ \bibnamefont
  {Back}}, \ and\ \bibinfo {author} {\bibfnamefont {T.}~\bibnamefont
  {Jungwirth}},\ }\href {\doibase 10.1103/RevModPhys.87.1213} {\bibfield
  {journal} {\bibinfo  {journal} {Rev. Mod. Phys.}\ }\textbf {\bibinfo {volume}
  {87}},\ \bibinfo {pages} {1213} (\bibinfo {year} {2015})}\BibitemShut
  {NoStop}%
\bibitem [{\citenamefont {Windsor}\ and\ \citenamefont
  {Stevenson}(1966)}]{windsor1966spin}%
  \BibitemOpen
  \bibfield  {author} {\bibinfo {author} {\bibfnamefont {C.}~\bibnamefont
  {Windsor}}\ and\ \bibinfo {author} {\bibfnamefont {R.}~\bibnamefont
  {Stevenson}},\ }\href@noop {} {\bibfield  {journal} {\bibinfo  {journal}
  {Proc. Phys. Soc. London}\ }\textbf {\bibinfo {volume} {87}},\ \bibinfo
  {pages} {501} (\bibinfo {year} {1966})}\BibitemShut {NoStop}%
\bibitem [{\citenamefont {Cheng}\ \emph {et~al.}(2015)\citenamefont {Cheng},
  \citenamefont {Xiao},\ and\ \citenamefont {Brataas}}]{cheng2015terahertz}%
  \BibitemOpen
  \bibfield  {author} {\bibinfo {author} {\bibfnamefont {R.}~\bibnamefont
  {Cheng}}, \bibinfo {author} {\bibfnamefont {D.}~\bibnamefont {Xiao}}, \ and\
  \bibinfo {author} {\bibfnamefont {A.}~\bibnamefont {Brataas}},\ }\href@noop
  {} {\bibfield  {journal} {\bibinfo  {journal} {arXiv preprint
  arXiv:1509.09229}\ } (\bibinfo {year} {2015})}\BibitemShut {NoStop}%
\bibitem [{\citenamefont {Eremenko}\ \emph {et~al.}(1968)\citenamefont
  {Eremenko}, \citenamefont {Novikov},\ and\ \citenamefont
  {Popkov}}]{eremenko1968anisotropy}%
  \BibitemOpen
  \bibfield  {author} {\bibinfo {author} {\bibfnamefont {V.}~\bibnamefont
  {Eremenko}}, \bibinfo {author} {\bibfnamefont {V.}~\bibnamefont {Novikov}}, \
  and\ \bibinfo {author} {\bibfnamefont {Y.~A.}\ \bibnamefont {Popkov}},\
  }\href@noop {} {\bibfield  {journal} {\bibinfo  {journal} {SOVIET PHYSICS
  JETP}\ }\textbf {\bibinfo {volume} {27}} (\bibinfo {year}
  {1968})}\BibitemShut {NoStop}%
\bibitem [{\citenamefont {Hulthen}(1936)}]{hulthen1936uber}%
  \BibitemOpen
  \bibfield  {author} {\bibinfo {author} {\bibfnamefont {L.}~\bibnamefont
  {Hulthen}},\ }\href@noop {} {\bibfield  {journal} {\bibinfo  {journal} {Proc.
  Akad. Wet. Amst.}\ }\textbf {\bibinfo {volume} {39}},\ \bibinfo {pages} {190}
  (\bibinfo {year} {1936})}\BibitemShut {NoStop}%
\bibitem [{\citenamefont {Kapelrud}\ and\ \citenamefont
  {Brataas}(2013)}]{kapelrud2013spin}%
  \BibitemOpen
  \bibfield  {author} {\bibinfo {author} {\bibfnamefont {A.}~\bibnamefont
  {Kapelrud}}\ and\ \bibinfo {author} {\bibfnamefont {A.}~\bibnamefont
  {Brataas}},\ }\href@noop {} {\bibfield  {journal} {\bibinfo  {journal} {Phys.
  Rev. Lett.}\ }\textbf {\bibinfo {volume} {111}},\ \bibinfo {pages} {097602}
  (\bibinfo {year} {2013})}\BibitemShut {NoStop}%
\bibitem [{\citenamefont {Cornelissen}\ \emph {et~al.}(2016)\citenamefont
  {Cornelissen}, \citenamefont {Peters}, \citenamefont {Duine}, \citenamefont
  {Bauer},\ and\ \citenamefont {van Wees}}]{cornelissen2016magnon}%
  \BibitemOpen
  \bibfield  {author} {\bibinfo {author} {\bibfnamefont {L.~J.}\ \bibnamefont
  {Cornelissen}}, \bibinfo {author} {\bibfnamefont {K.~J.}\ \bibnamefont
  {Peters}}, \bibinfo {author} {\bibfnamefont {R.~A.}\ \bibnamefont {Duine}},
  \bibinfo {author} {\bibfnamefont {G.~E.}\ \bibnamefont {Bauer}}, \ and\
  \bibinfo {author} {\bibfnamefont {B.~J.}\ \bibnamefont {van Wees}},\
  }\href@noop {} {\bibfield  {journal} {\bibinfo  {journal} {Phys. Rev. B}\
  }\textbf {\bibinfo {volume} {94}},\ \bibinfo {pages} {014412} (\bibinfo
  {year} {2016})}\BibitemShut {NoStop}%
\bibitem [{\citenamefont {Brataas}\ \emph {et~al.}(2015)\citenamefont
  {Brataas}, \citenamefont {Skarsv{\aa}g}, \citenamefont {Tveten},\ and\
  \citenamefont {Fj{\ae}rbu}}]{brataas2015heat}%
  \BibitemOpen
  \bibfield  {author} {\bibinfo {author} {\bibfnamefont {A.}~\bibnamefont
  {Brataas}}, \bibinfo {author} {\bibfnamefont {H.}~\bibnamefont
  {Skarsv{\aa}g}}, \bibinfo {author} {\bibfnamefont {E.~G.}\ \bibnamefont
  {Tveten}}, \ and\ \bibinfo {author} {\bibfnamefont {E.~L.}\ \bibnamefont
  {Fj{\ae}rbu}},\ }\href@noop {} {\bibfield  {journal} {\bibinfo  {journal}
  {Phys. Rev. B}\ }\textbf {\bibinfo {volume} {92}},\ \bibinfo {pages} {180414}
  (\bibinfo {year} {2015})}\BibitemShut {NoStop}%
\bibitem [{\citenamefont {Berger}(1996)}]{berger1996emission}%
  \BibitemOpen
  \bibfield  {author} {\bibinfo {author} {\bibfnamefont {L.}~\bibnamefont
  {Berger}},\ }\href@noop {} {\bibfield  {journal} {\bibinfo  {journal} {Phys.
  Rev. B}\ }\textbf {\bibinfo {volume} {54}},\ \bibinfo {pages} {9353}
  (\bibinfo {year} {1996})}\BibitemShut {NoStop}%
\bibitem [{\citenamefont {Wolfram}\ and\ \citenamefont
  {De~Wames}(1969)}]{wolfram1969surface}%
  \BibitemOpen
  \bibfield  {author} {\bibinfo {author} {\bibfnamefont {T.}~\bibnamefont
  {Wolfram}}\ and\ \bibinfo {author} {\bibfnamefont {R.}~\bibnamefont
  {De~Wames}},\ }\href@noop {} {\bibfield  {journal} {\bibinfo  {journal}
  {Phys. Rev.}\ }\textbf {\bibinfo {volume} {185}},\ \bibinfo {pages} {762}
  (\bibinfo {year} {1969})}\BibitemShut {NoStop}%
\bibitem [{\citenamefont {Cheng}(2014)}]{cheng2014aspects}%
  \BibitemOpen
  \bibfield  {author} {\bibinfo {author} {\bibfnamefont {R.}~\bibnamefont
  {Cheng}},\ }\emph {\bibinfo {title} {Aspects of antiferromagnetic
  spintronics}},\ \href@noop {} {Ph.D. thesis},\ \bibinfo  {school} {The
  University of Texas at Austin} (\bibinfo {year} {2014})\BibitemShut {NoStop}%
\end{thebibliography}
\end{document}